\documentclass[%
 reprint,
 amsmath,amssymb,
 aip, jcp
]{revtex4-2}
\usepackage[ruled]{algorithm2e}
\usepackage{mathdots}
\usepackage{mathrsfs}
\usepackage{graphicx}
\usepackage{dcolumn}
\usepackage{bm}
\usepackage{url}
\usepackage{amssymb}
\usepackage{amsthm}
\usepackage{float}
\usepackage{xcolor}
\usepackage[capitalise]{cleveref}
\usepackage{listings}

\SetCommentSty{mycommfont}

\definecolor{lightgray}{HTML}{F5F6FF}  
\definecolor{blue}{HTML}{3296DC}       
\definecolor{darkblue}{HTML}{3140B3}   
\definecolor{green}{HTML}{2CC486}      
\definecolor{background}{HTML}{EBECF7} 

\definecolor{support-info}{HTML}{0070C1}       
\definecolor{support-success}{HTML}{1B885C}    
\definecolor{support-undefined}{HTML}{8343D7}  

\definecolor{1}{HTML}{463480}
\definecolor{2}{HTML}{2f6c8e}
\definecolor{3}{HTML}{1e9c89}
\definecolor{4}{HTML}{5ec962}
\definecolor{5}{HTML}{dfe318}

\usepackage{tikz}
\usetikzlibrary{
    arrows,
    arrows.meta,
    positioning,
    shapes.geometric,
}

\tikzstyle{startstop} = [ellipse, minimum width=4cm, minimum height=1.5cm, text centered, fill=background!100, thick, text=black]
\tikzstyle{process} = [rectangle, fill=1!100, text=lightgray, align=center, minimum height=2cm]
\tikzstyle{decision} = [diamond, aspect=4, fill=4!100, text=lightgray, align=center, minimum height=2cm, inner sep=0pt]
\tikzstyle{databox} = [trapezium, trapezium right angle=100, trapezium left angle=80, fill=2!100, text=lightgray, align=center, inner xsep=0pt, minimum height=1.5cm, minimum width=5cm]
\tikzstyle{databox2} = [trapezium, trapezium right angle=100, trapezium left angle=80, fill=2!100, text=lightgray, align=center, inner xsep=0pt, minimum height=1.7cm, minimum width=6cm]
\tikzstyle{database} = [cylinder, shape border rotate=90, aspect=0.25, fill=2!100, text=lightgray, align=center, inner xsep=5pt, minimum height=3cm, draw=lightgray]

\tikzstyle{arrow} = [-{Latex[length=3mm]}, thick, draw=black]

\renewcommand{\r}{\mathbf{r}}

\usepackage{mathtools}
\DeclareMathOperator{\Tr}{Tr}

\begin{document}

\preprint{LA-UR}

\title{Machine-learned, finite temperature Fermi-operator expansions suitable for GPUs and AI-hardware}
\author{Stanislaw Kowalski}
\email{skowalski@lanl.gov}
\affiliation{Theoretical Division, Los Alamos National Laboratory}
\affiliation{Daniel Guggenheim School of Aerospace Engineering, Georgia Institute of Technology}
\author{Christian F. A. Negre}
\author{Anders M. N. Niklasson}
\author{Kipton Barros}
\email{kbarros@lanl.gov}
\author{Joshua Finkelstein}
\email{jdf@lanl.gov}
\affiliation{Theoretical Division, Los Alamos National Laboratory}

\date{\today}

\begin{abstract}
We present several finite-temperature recursive Fermi–operator expansion schemes based on the second-order spectral projection (SP2) method. Our approach builds on a previous observation that the electronic structure problem, as formulated through a recursive SP2 expansion, can be mapped onto the architecture of a deep neural network. Using this perspective, we generalize SP2 to finite electronic temperatures by constructing machine learning models that determine optimized recursive expansion coefficients. The same approach is also applied to the prediction of the electronic entropy for fractional occupation numbers. The coefficients are trained for a specified chemical potential and electronic temperature and are not available in closed analytical form. However, by employing an appropriate affine rescaling strategy to the Hamiltonian matrix, we eliminate the need to retrain the model for different temperatures and chemical potentials. Our approach avoids explicit diagonalization and relies solely on highly optimized matrix–matrix multiplication kernels. Compared to state-of-the-art diagonalization, we achieve an order-of-magnitude speedup in the single-particle finite-temperature density matrix calculation for small and moderately sized matrices on modern GPUs and dense matrix multiply units.
\end{abstract}

\maketitle

\section{Introduction}\label{sec:intro}

Quantum mechanics and electronic structure theory provide a framework for describing and predicting a wide range of phenomena and properties relevant to materials science, chemistry, and biology. In practice, however, the computational complexity often prohibits studies of relevant system sizes. 
In effective single-particle electronic structure theory, e.g. based on Hartree-Fock or Kohn-Sham (KS) density functional theory (DFT), \cite{Roothaan,hohen,KohnSham65} the computational cost is typically governed by an algebraic eigenvalue problem. 
Solving the eigenvalue problem through diagonalization exhibits a computational cost that scales cubically with the number of electrons, \cite{GGolub96} which for systems beyond a few thousand atoms typically becomes prohibitively expensive. 

Consequently, alternative approaches have been developed that are either lower complexity, such as numerically thresholded sparse matrix algebra \cite{SGoedecker99,Niklasson2011,DBowler12} or various divide-and-conquer-like approaches, \cite{WYang91,WYang95,KKitaura99,YNishimoto2014}, or have smaller cubic scaling prefactors \cite{JFinkelstein2021,JFinkelstein2022}. For local atomic-orbital-like basis-set representations, or semi-empirical approaches, some of the most efficient methods are based on recursive Fermi-operator expansion schemes, in which the effective single-particle density matrix is calculated through a recursive, dense or sparse matrix function expansion based on repeated matrix-matrix multiplications.\cite{APalser98,KNemeth00,AHolas01,ANiklasson02,ANiklassonAnomaly,DKJordan05,ERudberg11,PSuryanarayana13,EHRubensson14,Truflandier16} From the density matrix, we can then calculate, for example, electron densities, energies and forces directly, using matrix trace operations. 

The most efficient recursive Fermi-operator expansion schemes are restricted to zero electronic temperature, which results in an idempotent density matrix. These recursive schemes therefore cannot model systems with fractional orbital occupations, which are often needed to treat, for example, degenerate (or near degenerate) eigenstates at the chemical potential for chemical reactions or systems at elevated electronic temperatures. The purpose of this work is to demonstrate how various generalizations of the recursive Fermi-operator expansion scheme based on the second-order spectral projection (SP2) approach \cite{ANiklasson02} can be used to generate density matrices with fractional occupation at elevated electronic temperatures with a high level of accuracy. 

In previous generalizations of the recursive SP2 schemes to finite temperatures, the fractional occupation numbers were generated by a rescaled and truncated SP2 recursion scheme.\cite{SMniszewski2019} This approach can be used to calculate a density matrix with occupation numbers that resembles the distribution of the regular Fermi-function, but it is only approximate. Apart from this modified SP2 scheme, several alternative approaches for computing the finite-temperature density matrix without explicit diagonalization have also been proposed, including serial Chebyshev polynomial approximations \cite{Goedecker99} and recursive Padé expansions, \cite{ANiklassonAnomaly} though each method presents its own limitations. At low temperatures, Chebyshev expansions typically require very high polynomial orders to suppress Gibbs oscillations. While the introduction of smoothing kernels \cite{AWeisse06} can mitigate these oscillations, the resulting approximation still deviates from the exact Fermi function near the chemical potential. For small but finite temperatures, accurately resolving all states may require hundreds or even thousands of Chebyshev terms, with each term requiring a corresponding matrix multiplication. Although reduced-scaling techniques—where the number of matrix multiplications grows instead with the square root of the polynomial order—have been developed, \cite{JFinkelstein2023, MPaterson73, WLiang03} they may still fall short of practical efficiency and can exhibit ill-conditioning effects at large enough polynomial order.

Recursive Padé expansions avoid Gibbs oscillations and typically require only a small number of recursion steps, largely independent of temperature. Each recursion, however, requires the solution of a linear system, usually carried out with a few (1–4) conjugate gradient iterations, \cite{ANiklassonAnomaly} and while the number of recursions remains modest (6-8), the need to repeatedly perform linear solves increases the computational cost. Both Chebyshev- and Padé-based Fermi–operator expansions therefore incur a significant cost with a fairly large number of matrix multiplications or repeated linear solves. In practice, this cost can outweigh their benefit, making them less efficient than direct diagonalization, particularly on modern GPU architectures, where highly optimized vendor-supplied routines often achieve excellent performance.

Previous work has demonstrated highly-performant implementations of SP2 using GPUs \cite{Cawkwell_GPU_12,CBannwarth25} and, more recently, AI-oriented hardware such as Nvidia Tensor Cores, \cite{JFinkelstein2021, JFinkelstein2022} enabling evaluation of the zero-temperature density matrix in significantly less time than, for example, the optimized Nvidia cuSOLVER diagonalization routines across a wide range of matrix sizes. Here, it was also observed that the recursive algebraic structure of SP2 maps naturally onto the architecture of a deep neural network, suggesting that its recursive second-order form could be systematically exploited with machine learning. This viewpoint provides new flexibility in constructing recursive Fermi-operator expansions: rather than relying exclusively on analytically determined spectral projection (SP2) functions, we could introduce polynomial functions with machine learned coefficients (or, weights) at each iteration to capture the exact Fermi function at arbitrary temperature. This idea, introduced in Ref.\ \onlinecite{JFinkelstein2021}, forms the central focus of the present work and enables the reproduction of virtually exact Fermi functions at low computational cost for finite-temperature systems with fractional occupation numbers.

We begin by presenting some general theoretical background for density functional theory (DFT) and the SP2 method in Section \ref{sec:background}. In Section \ref{sec:algs} we discuss our generalizations of SP2 to finite temperatures based on several different types of neural network architectures, including details on how we train those models and how we can use them to efficiently compute the electronic entropy. In Section \ref{sec:workflow}, we demonstrate and provide a workflow for how these schemes can be implemented and applied to construct the finite-temperature density matrix. Lastly, in Section \ref{sec:numerics} we discuss our numerical results and computational performance using Nvidia Tensor Cores.

\section{Background} \label{sec:background}

\subsection{The quantum mechanical eigenvalue equation in KS-DFT}
In KS-DFT the effective single-particle electronic states (or molecular orbitals), 
\begin{equation}
\psi_i({\bf r}) = \sum_j v^{(i)}_j\phi_j({\bf r}), 
\end{equation}
are determined using a linear combination of local basis functions, $\{\phi_j\}$. Assuming an orthogonal basis set for simplicity, the expansion coefficients, $v^{(i)} = \{ v^{(i)}_j\}$, are given as the eigenvectors of the Kohn-Sham Hamiltonian matrix, 
\begin{equation}
H_{ij} = \langle \phi_i|{\widehat H}|\phi_j\rangle = \int \phi_i^\dagger({\bf r}){\widehat H} \phi_j({\bf r}) d{\bf r},
\end{equation}
i.e. from the matrix eigenvalue equation
\begin{equation} \label{KS_Eq}
Hv^{(i)} = \varepsilon_i v^{(i)}\;,
\end{equation}
where the set $\{\phi_j\}$ is orthonormal, i.e., where $\langle \phi_i|\phi_j\rangle = \delta_{ij}$. The electronic density is then given by
\begin{equation}\label{Density}
\rho({\bf r}) = 2\sum_i f_i |\psi_i({\bf r})|^2,
\end{equation}
with occupation factors $f_i$, which we can interpret as the probability of an electron residing in state $\psi_i$. The factor of 2 accounts for spin. In ground-state KS-DFT, the eigenstates have integer occupation numbers, i.e. \ $f_i = 1$ for the occupied states and $f_i = 0$ for the unoccupied ones and are characterized by a Fermion aufbau principle where the occupied
electronic states fill-in upwards from the lowest energy. These properties then lead to the occupation factor $f_i$ being represented by the Heaviside step function,
\begin{align}\label{eq:fi_step}
f_i = \Theta(\mu - \varepsilon_i)\;,
\end{align}
where $\mu$ is the chemical potential, or Fermi level, chosen such that such that the total sum of the occupation numbers equals the total number of occupied states, or equivalently, half the number of total electrons, i.e. 
\begin{equation}
\sum_i f_i = N_{\rm occ} = \frac{1}{2} N_\text{electrons}.
\end{equation}

In many cases, however, we need fractional occupation numbers $f_i$ that are different from 0 and 1; for example in metals, or when the electronic temperature is high in comparison to the electronic HOMO-LUMO gap, i.e. the difference in energies between the Highest Occupied Molecular Orbital and the Lowest Unoccupied Molecular Orbital. A solution is offered by extending ground-state KS-DFT to finite electronic temperatures, \cite{NMermin65,RParr89} where the electronic structure is treated in terms of an ensemble using fractional occupation numbers, where the $f_i$ are real numbers in the unit interval, i.e. $f_i \in [0,1]$ for every $i$. At low lying states the occupation numbers are close to 1 and for states at high energies the occupation numbers are close to 0. However, for intermediate states, with energies close to $\mu$, which normally separates the fully occupied states from the unoccupied, the occupation number varies continuously between 1 and 0, and how fast it changes depends directly on the temperature.

In finite temperature KS-DFT, \cite{RParr89} the function that describes the fractional occupation numbers is the Fermi function
\begin{equation}\label{FD_Func}
f_i \equiv f(\varepsilon_i)= \left[e^{\beta(\varepsilon_i - \mu)}+1\right]^{-1}\;,
\end{equation}
where $\varepsilon_i$ is, again, the $i$-th eigenenergy, $\beta = 1/(k_B T)$ is the inverse temperature, and $k_B$ is Boltzmann's constant. At elevated temperatures the step-like Fermi function becomes increasingly smeared, whereas in the low-temperature limit, i.e. when $\beta \rightarrow \infty$, the Fermi function in Eq.~(\ref{FD_Func}) becomes exactly the step function in Eq.~(\ref{eq:fi_step}).

Evaluating the Fermi function of the KS Hamiltonian matrix, $H$, gives the density matrix, $D$, where
\begin{equation}
D = \left[e^{\beta(H - \mu I)}+I\right]^{-1}, 
\end{equation}
from which the expectation value of any quantum observable $A$ can be calculated as
\begin{equation} \begin{split} \label{eq:observables}
\langle A \rangle &= 2\Tr(DA) \;,
\end{split} \end{equation}
and, for example, the electron density, $n({\bf r})$, is given by
\begin{equation}
n({\bf r}) = 2\sum_{ij} D_{ij} \phi_i({\bf r}) \phi_j({\bf r})\;,
\end{equation}
where the factor $2$ has been included to account for spins.

\subsection{Second-order spectral projection (SP2)}
Construction of an analytical quadratically convergent method for the zero-temperature step function can be derived by designing an approach that iteratively projects the higher eigenenergies of the KS Hamiltonian to 0 and the lower eigenenergies to 1 in order to achieve a desired occupation number, corresponding to a purified idempotent density matrix. The original SP2 scheme \cite{ANiklasson02,EHRubensson14,ERudberg11}  starts with a linear transformation, $X_0$, of the Hamiltonian matrix $H$, where the spectrum of $H$ is mapped to the unit interval $[0,1]$ in reverse order (we refer to this as the initial spectral flip). Iterative spectral projection updates, $X_{i+1} = F(X_i)$, are then applied with second-order matrix polynomials that bring the eigenvalues closer to either 0 or 1, sometimes referred to as {\it purification}. A squaring operation, $F(X) = X^2$, reduces all eigenvalues towards 0, and a similar attractor can be constructed around 1 with $F(X)= I-(I-X)^2 = 2X-X^2$, which brings eigenvalues closer to 1. At the same time, this process raises or lowers the total trace of the system, which can be used to successively approach the desired occupation number at convergence. The SP2 scheme is therefore sometimes also referred to as a trace-correcting scheme.

We now describe the SP2 method more precisely. Given a Hamiltonian matrix, $H$, and spectral bounds $\varepsilon_\text{min} \leq \lambda_H \leq  \varepsilon_\text{max}$, $H$ is first transformed so that its spectrum is flipped and lies inside the unit interval [0,1], i.e.\
\begin{align}\label{eq:sp2-spectrum_flip}
     H' \equiv \frac{1}{\varepsilon_\text{max}-\varepsilon_\text{min}}(\varepsilon_\text{max}I-H)\;.
\end{align}
Setting $X_0 = H'$, an SP2 projection is then given by
\begin{align} \label{eq:sp2}
    X_{i+1} &= \left\{ \begin{array}{ll}
        X_i^2 & \text{if } |{\rm Tr}(X_i^2) - N_{occ}|  \\
        & < |{\rm Tr}(2X_i-X_i^2) - N_{occ}| \\
        2X_i - X_i^2& \text{otherwise} 
        \end{array} \right. 
\end{align}
with the algorithm terminating once $|\Tr(X_{i+1})-N_{\rm occ}|$ is sufficiently small or depending on the available numerical accuracy. \cite{AKruchinina16} Here the polynomial projection in each iteration, either $X^2$ or $2X-X^2$, is chosen to be the one that brings the trace of the applied matrix polynomial closest to $N_{\rm occ}$. Other choices are possible, including pre-determined sequences of the spectral projections.

If the chemical potential $\mu$ is known instead of the occupation, Eq.~(\ref{eq:sp2}) can be adapted. A similar update rule can be used with the updates of $\mu$ to determine which polynomial is chosen instead of a trace condition. Putting
\begin{equation}
\mu_0=\mu'= \frac{1}{\varepsilon_\text{max}-\varepsilon_\text{min}}(\varepsilon_\text{max}-\mu)\;,
\end{equation}
and again,
\begin{equation}
X_0=H'\equiv \frac{1}{\varepsilon_\text{max}-\varepsilon_\text{min}}(\varepsilon_\text{max}I-H)\;,
\end{equation}
iteration $i+1$ of SP2 for a known chemical potential $\mu$ and unknown occupation is
\begin{equation} \begin{split}\label{eq:sp2-mu}
    X_0 &= H'\; \\
    X_{i+1} &= \left\{ \begin{array}{ll}
        X_i^2 & \text{if } |\mu_i^2 - \mu'| \\
        & < |2\mu_i - \mu_i^2 - \mu'| \\
        2X_i - X_i^2& \text{otherwise} 
        \end{array} \right. \\
\end{split}\end{equation}
where the updates to $\mu'$ are tracked and $\mu_{i+1}$ is chosen such that
\begin{equation}
    \mu_{i+1} = \left\{ \begin{array}{ll}
        \mu_i^2 &   \text{if } |\mu_i^2 - \mu'| \\
        & < |2\mu_i - \mu_i^2 - \mu'| \\
        2\mu_i - \mu_i^2& \text{otherwise} 
        \end{array} \right. \\
\end{equation}
Similarly to Eq.~(\ref{eq:sp2}), this method can be terminated if $|\mu_{i+1} - \mu'|$ is small enough relative to a preselected tolerance. In other words, we choose the polynomial $X^2$ at iteration $i+1$ if $\mu_{i}^2$ is closest to $\mu'$ and choose $2X-X^2$ otherwise. Interestingly, since $\mu' \in (0,1)$ is essentially arbitrary, the procedure in Eq.~(\ref{eq:sp2-mu}) in fact shows how any real number in the unit interval can be approximated with a second-order recursive polynomial expansion. Crucially, the function applied (i.e. the precise sequence of either $X_i^2$ or $2X_i - X_i^2$) is now dependent strictly on the parameter $\mu$, and entirely independent of the input $X$. 

Before reaching full convergence and idempotency, the SP2 scheme generates a smooth approximation to a step function. Because the projection polynomials are continuously increasing functions, no wiggles or Gibbs-like oscillations appear. A truncated SP2 scheme can therefore be used to capture the effect of fractional occupation numbers at elevated electronic temperatures.\cite{SMniszewski2019}  With additional rescaling, this truncated version of SP2 can capture the slope of the Fermi function at any chosen temperature. However, this approach does not always yield a sufficiently accurate Fermi-operator expansion at elevated temperatures. The goal of this paper is to generalize the recursive SP2 expansion to provide more exact and systematically improvable approximations to the Fermi operator.

\section{Algorithms} \label{sec:algs}

The recursive SP2 Fermi-operator expansion scheme converges to a matrix step function of the Hamiltonian, with the step formed at the chemical potential. This corresponds to a matrix Fermi function at zero electronic temperature, $T=0$, where the occupation numbers, $f_i$, are either 0 or 1. At elevated electronic temperatures this is no longer the case and the fully converged SP2 scheme is no longer valid. However, by re-interpreting and generalizing the SP2 scheme in terms of various deep neural network architectures, with flexible weights and bias coefficients, we can train or optimize these coefficients such that the different expansions approximate the Fermi functions with fractional occupation numbers at any chosen electronic temperature and chemical potential. 

First we will discuss a deep neural network formulation of the SP2 scheme itself and then propose a set of extensions and generalizations of the SP2 scheme for finite temperature Fermi-operator expansions. We then apply the recursive approach also to the approximation of the electronic entropy function, which is needed to estimate the electronic free energy in the case of fractional occupation numbers. 

The method that we use for the training or optimization of the flexible expansion coefficients is presented afterwards. This optimization is a non-trivial task and requires an accurate initial guess. The initial guess is provided by the original SP2 scheme or its non-converged, truncated version, which resembles the shape of a Fermi function. For the approximation of the entropy function we can use a scaled polynomial of the Fermi function as an accurate initial guess. The training and optimization of the schemes is here discussed for energies $x \in [0,1]$. For expansions using Hamiltonians with other energy ranges, an additional linear transform of the spectrum to the unit interval, discussed in Sec. \ref{ssec:rescaling}, would then be needed.

\subsection{Deep Spectral Refinement Networks}

The recursive SP2 scheme in Eq.~(\ref{eq:sp2}) looks very similar to the computational structure of a deep neural network (DNN), and this observation turns out to be quite powerful. \cite{JFinkelstein2021} In Ref.~\onlinecite{JFinkelstein2021}, the SP2 method described in Eq.~(\ref{eq:sp2}) was expressed as a DNN using weights $W_i$, biases $B_i$, network layers $X_i$, and a non-linear activation function $h(S_i) = S_i^2$. In each recursion or DNN layer,
\begin{align} \label{eq:S_i}
    X_{i+1}=h(S_i)=(W_iX_i+B_i)^2\;,
\end{align}
where
\begin{align}
    W_i &= \sigma_i I, \\
    B_i &= (1-\sigma_i)S_{i-1}, \\
     S_i &= W_i X_i+B_i \;.
\end{align}
Here $\sigma_i=\pm 1$, and the sign can be chosen based on, for example, the trace of $S_i$, such that the desired occupation is reached at convergence.

\begin{figure}
    \centering
    \includegraphics[width=\linewidth]{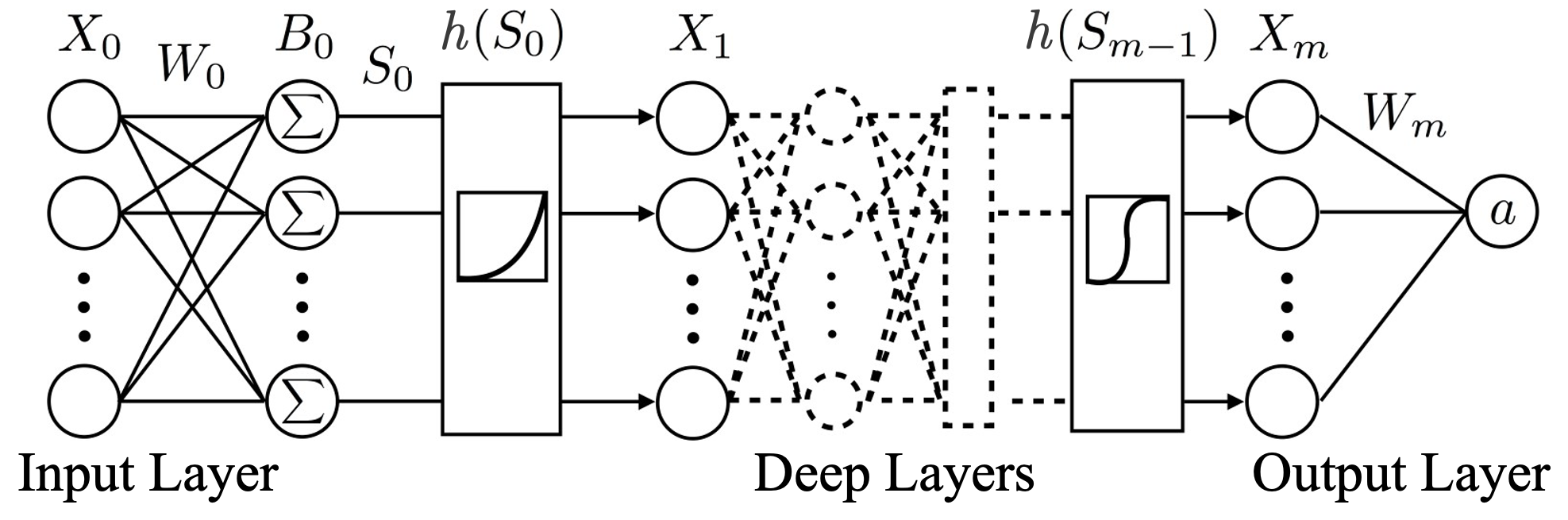}
    \caption{Original description of DNN-SP2 in Ref.~\onlinecite{JFinkelstein2021}, showing weight $W_i$ and bias $B_i$ applied to the matrix $X_i$ at each layer before squaring, which serves as the nonlinear activation function $h$. This form is equivalent to the MLSP2 method described in Eq.~(\ref{eq:mlsp2_alt}), without a running accumulator, and expresses the classic SP2 algorithm.}
    \label{fig:MLSP2 structure}
\end{figure}

The matrix $S_i$, which is obtained from applying the weight and bias to the previous network layer, is the density matrix approximation at layer $i$. Notice that choosing $\sigma_i=1$ leads to 
\begin{align}
S_{i} &= 1S_{i-1}^2 + (1-1)S_{i-1} = S_{i-1}^2 \;,
\end{align}
and choosing $\sigma_i=-1$ gives
\begin{align}
S_{i} &= -1S_{i-1}^2 + (1+1)S_{i-1} = 2S_{i-1} - S_{i-1}^2\;,
\end{align}
which is equivalent to the application of the polynomials $X^2$ and $2X-X^2$ in Eq.~(\ref{eq:sp2}) at the $i$-th step. We call this DNN formulation of SP2, DNN-SP2. \cite{JFinkelstein2021} The neural network structure of DNN-SP2 is schematically shown in Fig.~\ref{fig:MLSP2 structure}, which also includes a refinement activation function in the last layer that is different from $S^2$. By viewing the recursive Fermi-operator expansion from the perspective of a neural network that predicts an outcome given some input data, it is natural to consider how we can optimize or train the network for various predictions. In our original study we modified the coefficients in each iteration to accelerate the convergence towards a step function determining the density matrix at zero temperature. This was achieved by using prior information about HOMO-LUMO eigenvalues \cite{JFinkelstein2021}. Here we will instead optimize the coefficients to generate an extremely close approximation of the exact Fermi-function expansion at elevated electronic temperatures.

Notably, the input and nonlinearity of the DNN-SP2 scheme have a slightly different structure compared to a usual neural network. Although the input is part of a vector space, it has the additional structure of a matrix, which is preserved by the linear layers and, crucially, the nonlinear activation function, both of which preserve the eigenspaces of the matrix, and act only on the eigenvalues (doing so by multiplication, without having to calculate an eigendecomposition). In this sense, the network can be described as a scalar function acting on the eigenvalue spectrum. When we optimize the coefficients of the DNN network we can thus do so for a scalar function, which then can be applied in the recursive DNN matrix expansion, where we assume all matrices are symmetric or Hermitian.

The machine-learning optimization of the expansion coefficients in the network can be seen as a way to approximate an analytic function. However, instead of using a traditional approximation based on, for example, orthogonal polynomials, we use a recursive expansion. In this way, we can reach much higher polynomial orders in fewer steps, potentially achieving a more accurate expansion at lower cost. Alternatively, we may view the expansion as a deep neural operator that can be trained to predict the desired results. \cite{JFinkelstein2021} We find that both viewpoints contribute to the analysis and results, a detailed description of which, including a comparison to state of the art architectures such as group-Equivariant Neural Networks, can be found in Appendix \ref{appendix:Analysis of Spectral Refinement Networks}.

\subsubsection{Quadratic Layers with Accumulation (MLSP2)} \label{ssec:MLSP2}
In the same way that SP2 operates on the eigenvalue spectrum of a symmetric or Hermitian matrix rescaled to $[0,1]$, our machine-learned DNN models can be trained on real values $x \in [0,1]$. Our first generalization of SP2, called machine-learned SP2 (MLSP2), was first proposed in Ref.\ \onlinecite{JFinkelstein2021} and is given by a sequence of second-order projection neural network layers. For $0 \le i < n$, we define layer $i+1$ as
\begin{equation} \begin{split} \label{eq:mlsp2}
    x_{i+1} &= \theta_{i,3} + \theta_{i,2} \; x_i + \theta_{i,1} \; x_i^2 \\
    A_{i+1} &= A_i + \theta_{i,4}\; x_i 
\end{split}  \end{equation}
with the Fermi function $f$ approximated by the final output layer,
\begin{equation} \begin{split} \label{eq:mlsp2-output} 
    f(x) \approx A_n + x_n \; .
\end{split} \end{equation}
The zeroth layer, $x_0$, emulates the spectrum flip of SP2, and we define it to be $x_0 = 1-x$. 

Without the accumulator, Eq.~(\ref{eq:mlsp2}) reduces to the SP2 scheme when $\theta_{i,1}=\sigma_i$, $\theta_{i,2}=(1-\sigma_i)$ and $\theta_{i,3}=0$. The method shown in Eq.~(\ref{eq:mlsp2}) is however 
more flexible compared to the original SP2 expansion. In particular, instead of approximating the Fermi function with a scaled and truncated SP2 scheme, 
\cite{SMniszewski2019} the coefficients in the MLSP2 scheme can be trained to fit the true Fermi function at any given temperature. 

In each layer of the MLSP2 scheme there are four coefficients that can be optimized. For the training, we use a non-linear optimization procedure based on the Levenberg-Marquardt approach discussed in Section \ref{sec:training}, with the exact Fermi function as the target. While this is an efficient method, our optimization is highly non-linear and requires a good initial guess. Because the MLSP2 scheme can be seen as a direct generalization of the SP2 scheme, the coefficients of the original scaled or truncated SP2 scheme \cite{SMniszewski2019} can be used as an initial guess.

The accumulator, $A_i$, is a memory- and arithmetically-efficient way to provide more flexibility (and tunability) of the final output prediction with respect to each layer in a manner that is not wholly dependent on the non-linear interaction with all subsequent layers, smoothing the loss landscape \cite{Hao2018} and improving accuracy. A pseudocode for the MLSP2 scheme as presented in Eq.~(\ref{eq:mlsp2}) for energies in the unit interval is shown in Alg.~\ref{alg:mlsp2-1d}.

\begin{algorithm}
\caption{in-place MLSP2 approximating the Fermi function, $f(x)$, for $x\in [0,1]$} \label{alg:mlsp2-1d}
$x = 1 - x$ \;
$A = 0$ \;
\For{$0\leq i \leq n-1$}
{
$a,b,c,d = \theta_{i,1}, \theta_{i,2}, \theta_{i,3}, \theta_{i,4}$ \;

$A = A + d \; x$ \;

$x = ax^2 + bx + c$ \;
}
$f = A + x$\;
\end{algorithm}

Eq.~(\ref{eq:mlsp2}) can naively be thought of as having $4n$ parameters, but strictly speaking this model has $2n+2$ degrees of freedom, as it can be re-expressed in a DNN-SP2-like form,
\begin{equation} \begin{split} \label{eq:mlsp2_alt}
     x_{i+1} &= (   \theta_{i,1} + x_i )^2 \\
     A_{i+1} &=  A_i +  \theta_{i,2}\;  x_i 
\end{split}  \end{equation}
with final output after $n$ layers given by
\begin{equation} \begin{split} \label{eq:mlsp2_alt-output}
    f(x) \approx A_n + \theta_{n,1} + \theta_{n,2} \; x_n
\end{split} \;. \end{equation}
Here, the weights $\{\theta\}$ need to be adjusted from Eq.~(\ref{eq:mlsp2}) and the equivalency between these two forms is shown in Appendix \ref{appendix:embeddings}. This formulation may be more optimal on some hardware architectures from an implementation perspective since it requires fewer matrix operations in each iteration.

\begin{figure}
    \centering
    \includegraphics[width=\linewidth]{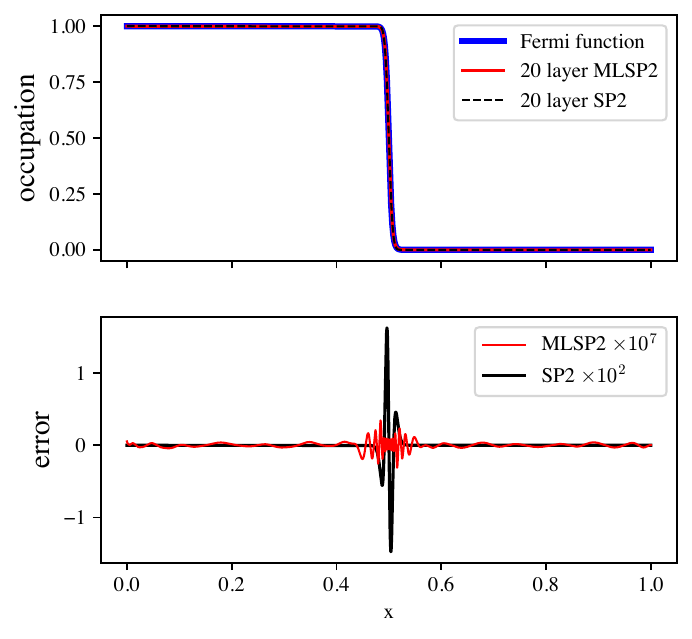}
    \caption{(top) Comparison of a scaled and truncated SP2 approximation of the Fermi function as described in Ref.~\onlinecite{SMniszewski2019} with $\beta=308.3$ and $\mu=1/2$ with 20 layers and a trained MLSP2 model from Eq.~(\ref{eq:mlsp2}) for $x \in [0,1]$. (bottom) Here we show the error measured pointwise between each approximation with the true Fermi function $f$. The SP2 error is on the order $10^{-2}$, and the MLSP2 error is on the order of $10^{-7}$. The errors are magnified by these respective factors to make them more visible.}
    \label{fig:sp2 vs mlsp2}
\end{figure}

In the upper panel of Fig.~\ref{fig:sp2 vs mlsp2} we show the approximations to the Fermi function inside the unit interval, with $\beta=308.3$ and $\mu=1/2$, using both the scaled and truncated SP2 expansion \cite{SMniszewski2019} and the optimized MLSP2 scheme -- both using 20 layers. In the lower panel we show the corresponding scaled pointwise errors. The SP2-based approximation has a max error of about $0.01$, while the MLSP2 scheme is correct to within $10^{-7}$.

\subsubsection{Maximally Expressive Quadratic (Max-SP2)}

We can generalize the SP2 scheme in Eq.~(\ref{eq:mlsp2_alt}) or Eq.\ (\ref{eq:S_i}) by allowing a contribution from not only the previous layer, but also every layer before it. We call this scheme Max-SP2. Every layer is the square of a linear combination of all previous layers, which introduces the idea of ``skip connections" from machine learning. \cite{Huang2017} Again, we define the zeroth layer as $x_0 = 1-x$ for $x \in [0,1]$ in order to capture the SP2 spectrum flip. 

For $0 \le i < n$, the update rule for layer $i+1$ is then
\begin{equation} \begin{split} \label{eq:maxsp2}
x_{i+1} &= \left( \delta_i + \sum_{j=0}^i \theta_{i,j} \; x_j  \right)^2, \\
A_{i+1} &= A_i + \gamma_i \; x_i 
\end{split}  \end{equation} 
with the final output layer yielding the Max-SP2 approximation of the Fermi function, 
\begin{align}
    f(x) \approx A_n + \gamma_n x_n \;.
\end{align}
Here $\{\theta\}$, $\{\gamma\}$ and $\{\delta\}$ are trainable parameters scaling the various layer connections and the accumulation into the final output. The parameter $\gamma_n=\pm 1$ is determined by whether, in the SP2 initialization, $x_n = x_{n-1}^2$ or $x_n = 2x_{n-1} -x_{n-1}^2$, respectively. A pseudocode for the Max-SP2 scheme for energies trained on the unit interval is presented in Alg.\ \ref{alg:maxsp2-1d}. 

The Max-SP2 approach in Eq.~(\ref{eq:maxsp2}) is quite flexible and can achieve higher accuracy with far fewer layers than the previous two Fermi-function approximations. The Max-SP2 scheme has $\mathcal O(n^2)$ parameters to tune which provides flexibility, but requires more memory, since it needs to store all previous matrices -- one for each layer -- in memory. By including the information $\{x_i\}$, from all previous layers, this formulation provides a maximally expressive network architecture that uses a squaring as its nonlinear activation function and provides a much larger number of parameters (per multiplication) to tune at the cost of additional matrix additions. Section \ref{sec:numerics} explores this tradeoff, and empirically finds it largely in favor of MLSP2 at high $\beta$ in single-precision arithmetic, although this Max-SP2 architecture could be necessary for double precision.

\begin{algorithm}
    \caption{Max-SP2 approximating $f(x)$ for $x \in [0,1]$}\label{alg:maxsp2-1d}
    $x_0 = 1-x$ \;
    \For{$0 \leq i \leq n-1$}{
        $x_{i+1} = \delta_{i} $ \;
        \For{$0 \leq j \leq i$}{
            $x_{i+1} = x_{i+1} + \theta_{i,j}x_j$ \;
        }
        $x_{i+1} = (x_{i+1})^2$ \;
    }
    $f = \gamma_0$ \;
    \For{$1 \leq i \leq n$}{
        $f = f + \gamma_i x_i$ \;
    }
\end{algorithm}

Further flexibility can be achieved by generalizing the $x^2$ activation function and replacing it with an $x \times y$-like activation function, which still preserves matrix symmetry if both $x$ and $y$ are arbitrary polynomials in the same original matrix (which is guaranteed by the structure of our network).  
This approach provides a more expressive and flexible alternative to Eq.~(\ref{eq:maxsp2}) where layer $i+1$ is given by
\begin{equation} \begin{split} \label{eq:arbitrary}
x_{i+1} &= \left( \delta_i + \sum_{j=0}^i \phi_{i,j} \; x_j \right) \left( \delta_i' +\sum_{j=0}^i \psi_{i,j} \; x_j \right), \\
A_{i+1} &= A_i + \gamma_i \; x_i \;.
\end{split} \end{equation} 
for $x\in [0,1]$ and $0 \le i < n$. Again, the zeroth-layer emulates the SP2 spectrum flip, and the output at the last layer approximates the target Fermi function
\[
f(x) \approx A_n + \gamma_n \; x_n \;.
\]
In principle, this version nearly doubles the number of parameters compared to  Eq.~(\ref{eq:maxsp2}), providing further flexibility in the Fermi-function approximation.
We refer to this architecture as Arbitrary multiplication SP2 (Arb-SP2), but it is not explored further in this work.

\subsubsection{Recombinant Layers and Accumulators (Skip-SP2)}

We can construct a ``compressed" form of Max-SP2, which we call Skip-SP2, by modifying Eq.~(\ref{eq:maxsp2}) in order to limit the use of memory. This method further leverages the idea of skip connections where now each layer $i$ is the square (or, activation function) of a linear combination of the $k$ most recent layers, as well as $K$ running accumulators and a constant offset $\alpha_{n,k}$. 

With the zeroth layer defined as $x_0=1-x$, layer $i+1$ of Skip-SP2 is
\begin{equation} \begin{split}
    x_{i+1} &= \left( \alpha_{i,k} +  \sum_{j=0}^{k-1} \alpha_{i,j} \; x_{i-j}  + \sum_{j=1}^K \beta_{i,j} A_{i,j} \right)^2, \\
    A_{i+1,j} &= A_{i,j} + \gamma_{i,j} \; x_i\;,
\end{split} \end{equation}
for $x \in [0,1]$ and $0 \le i < n$. At $i=n$, the output layer
\begin{align}
f(x) \approx x_n + \sum_{j=1}^K A_{n,j}
\end{align}
approximates the Fermi function at $x$. Here $k$ is number of layers stored, $K$ is the number of accumulators, and $\{ \alpha_{i,j} \}, \{ \beta_{i,j}\}, \{\gamma_{i,j}\}$ are the trainable coefficients of the linear recombination of the skip connections, accumulators contributing to $x_i$, and $x_i$ contributing to the accumulators. 

This model captures the full expressibility of quadratics; any MLSP2 scheme can be embedded in Skip-SP2, but Skip-SP2 only has $(1+k+2K)n+2$ parameters, allowing for deeper fine-tuning while keeping memory constant in layer count $n$. Just one skip connection, $k=1$, allows for qualitatively different behavior such as fusing two layers into an arbitrary 4th-order polynomial. Even not accounting for potential emergent behavior, this enables, for example, initialization with a 4th order spectral projection, as found in Ref.~\onlinecite{ANiklasson02}.

\subsection{Approximate Entropy Function}
At elevated electronic temperatures we need to be able to calculate the electronic free energy, 
\begin{equation}
    G \equiv U - \mu N_\text{occ} - TS \;,
\end{equation}
which includes the internal energy, $U = \langle H \rangle =  2\Tr[DH]$ by Eq.~(\ref{eq:observables}), and the entropy contribution, $S$, from fractional occupation numbers. Since $N_{\rm occ} = 2\Tr[D]$, the free energy can be expressed in the matrix form 
\begin{equation}
    G = 2\Tr[D(H-\mu I)] - 2TS \;,
\end{equation}
where the entropy is given by
\begin{equation}\label{eq:entropy-matrix}
    S = - k_B \Tr [ D \ln D + (I-D) \ln (I-D) ],
\end{equation}
with the factor of 2 included to account for spin. 

Directly computing the entropy from Eq.~(\ref{eq:entropy-matrix}) is generally challenging, as it requires evaluating matrix logarithms. We therefore adopt an alternative approach that treats the free energy as a function of the energy itself. Indeed, the free energy can be written as the trace $G = 2\Tr[g(H)]$ of the free energy function, 
\begin{equation} \label{eq:free_energy}
    g(x) = f(x)(x-\mu) - \beta^{-1} s(x) \;,
\end{equation}
where $x$ corresponds to an energy eigenvalue of $H$, and where $s: \mathbb{R} \to [0,\infty)$ is the entropy function given by
\begin{equation} \begin{split} \label{eq:entropy}
    s(x) &= -f(x) \ln f(x) - (1-f(x))\ln(1-f(x)) \;.
\end{split} \end{equation}
If $f(x)$ and $s(x)$ are already known, then computing Eq.~(\ref{eq:free_energy}) is much more straightforward than computing $s$ as a function of $f$ itself 
\begin{align}
s(f) &= -f \ln f - (1-f) \ln (1-f) \; ,
\end{align}
as in Eq.~(\ref{eq:entropy-matrix}).

Previous approaches to approximate the electronic entropy for recursive Fermi-operator expansion schemes have used either a variationally-exact entropy computed by integrating the inverse of the SP2 scheme on the eigenvalues closest to the Fermi level (estimated via Lanczos method)\cite{SMniszewski2019} or polynomial expansions of the fractional occupation numbers or the density matrix directly.\cite{ANiklasson15} The former approach cannot be used here, as the individual MLSP2 projections are not formally one-to-one on $[0,1]$, and each iteration therefore does not have a well-defined inverse, over which integration would need be performed for the exact entropy calculation. In the latter case of a polynomial expansion in fractional occupation, $s$ was approximated via
\begin{equation}\label{eq:approx-s}
    s(f) \approx  \sum_{i = 1}^m c^{(m)}_{i} f^i(1-f)^i ,
\end{equation}
where the coefficients, $\{ c_i^{(m)} \}$, for various expansion orders $m$ are given in Ref.\ \onlinecite{ANiklasson15}, (though we note here that the coefficients with even order in $i$ have a sign error and should have opposite sign). The accuracy is good only around the chemical potential at $f = 0.5$ and quite poor towards the end points, although exact at $f = 0$ and $f = 1$. Unfortunately, this will always be the case for any polynomial expansion of $s(f)$, as $s(f)$ has unbounded derivatives at $f=0$ and $f=1$ due to the logarithms. 

To get around this, we expand the entropy, $s$, as a function of the energy, $x$, i.e., as  $s(x)$ instead of as a function of the occupation number, $s(f)$. This can be achieved via a three-step procedure. First, we rescale the energy as $y = \alpha(x-\mu) + \mu$ for some optimal parameter $\alpha$. Then, we calculate $f(y)$ (using for example MLSP2), and lastly, we calculate $s(y(x))$, using the approximation in Eq.~(\ref{eq:approx-s}) for $m = 1$.  The parameter $\alpha$ can be optimized in various ways, for example, we may choose $\alpha$ such that $s''(\mu)$ is exact or such that the global max error of the entropy approximated by $(4\ln 2) f(y) (1 - f(y))$ for $x \in \mathbb R$ is minimized. The first case leads to $\alpha = 1/\sqrt{(2\ln 2)}$ and the second case to $\alpha \approx 0.842704$. This simple approach provides a surprisingly good estimate of the entropy function $s(x)$ over the real line, and, as we will show later, can be used as an initial guess in a non-linear optimization scheme for the entropy.

\begin{figure}
    \centering
    \includegraphics[width=\linewidth]{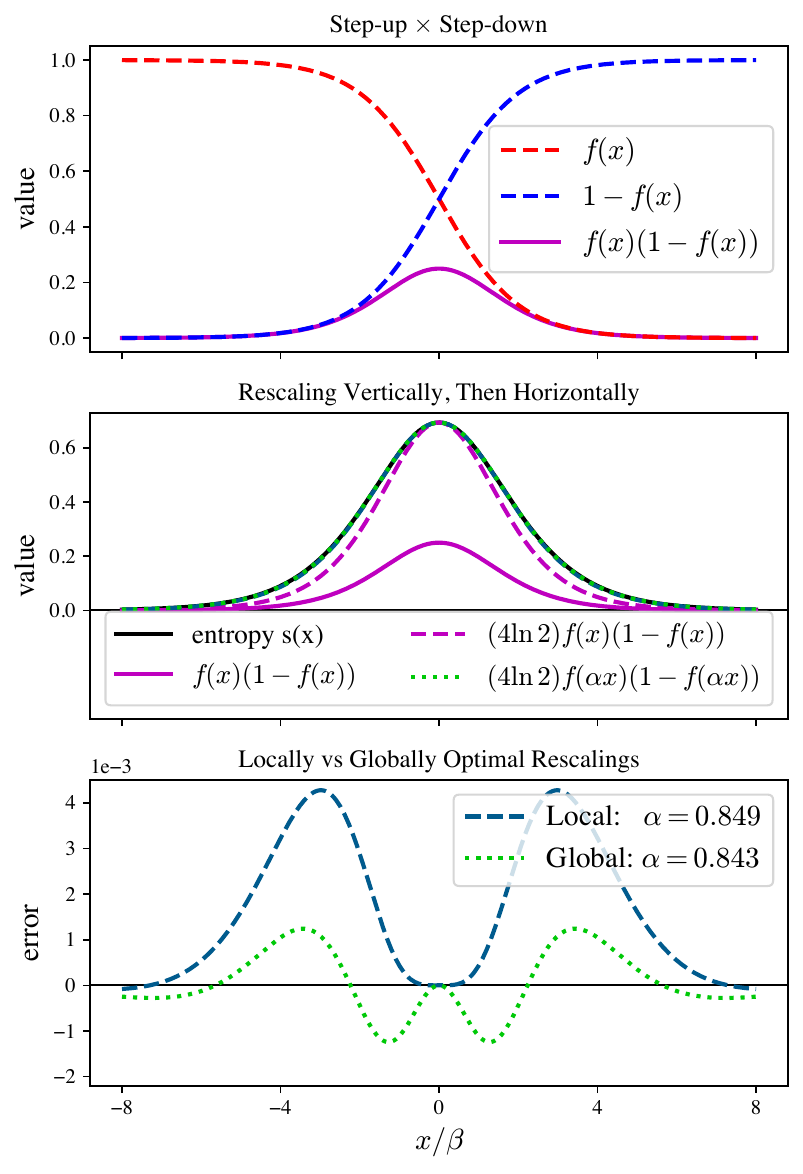}
    \caption{A graphical process for creating approximations to $s(x)$ that serve as excellent weight initializations, which accelerates training dramatically. The 3rd panel displays the pointwise error of the two best approximations shown in the 2nd panel. The chemical potential, $\mu$, has been set to zero for simplicity of display.}
    \label{fig:entropy accuracy}
\end{figure}

Figure \ref{fig:entropy accuracy} illustrates the construction and behavior of the $\alpha$-rescaled entropy ansatz for $m=1$ in Eq.~(\ref{eq:approx-s})
as the product of a ``step-down" function, $f(x)$, with a ``step-up" function, $1-f(x)$, resulting in a Gaussian-like function visually similar to $s(x)$. The first panel shows how the product of this step-up and step-down looks. The second panel then shows the rescaling of this resultant product vertically to match the peak of $s(x)$, and then horizontally to match the curvature of $s(x)$ at that peak. The third panel shows an error comparison between $\alpha$ set to the locally optimal value matching the second derivative, $\alpha = \frac{1}{\sqrt{2 \ln 2}} \approx 0.849322$, or to the value, $\alpha \approx 0.842704$, given by a global max error optimization over the range of $f$, the unit interval.

With this three-step approximation to $s(x)$, we can now develop a recursive, neural network entropy method based on the MLSP2 scheme. We assume all energies are already normalized, i.e. $x ,\mu \in[0,1]$, and define layer zero and layer $i+1$ as
\begin{equation} \begin{split} \label{eq:entropy expansion}
    x_0 &= \alpha (x - \mu) + \mu, \\
    x_{i+1} &= \theta_{i,1} \; x_i^2 + \theta_{i,2} \; x_i + \theta_{i,3}, \\
    A_{i+1} &= A_i + \theta_{i,4} \; x_i, 
\end{split} \end{equation}
for $0\le i < n$. The final layer, defined using the $m=1$ approximation in Eq.(\ref{eq:approx-s}), approximates the entropy $s$ as
\begin{align}\label{eq:entropy expansion-s}
s(x) \approx (4\ln 2)(A_n + x_n)(1-(A_n + x_n))\;.
\end{align}
This recursive expansion for $s(x)$ can then be trained in the same manner as the MLSP2 scheme with the difference being that now we have an $\alpha$-rescaled $x$ as input and the $m = 1$ approximation as the final layer. In Eq.~(\ref{eq:entropy expansion}) it is also worth noting that we require $|\alpha|\le 1$ be chosen so that $x_0$ maps $[0,1]$ to $[0,1]$. As opposed to the pure MLSP2 method, we do not need to implement the zeroth-layer flip $x_0=1-x$ since the entropy is symmetric about $\mu$.
 
From this recursive expansion, we are lead to two straightforward options. 1) we train the entropy model by using a previously trained MLSP2 as an initial guess, and then generate a new set of model coefficients. Or 2) we simply use a previously trained MLSP2 model with Eq.~(\ref{eq:entropy expansion}), and then plug this result directly into the final layer in Eq.~(\ref{eq:entropy expansion-s}). We denote these entropy model coefficients as $\{\theta_{\rm entropy}\}$ and $\{\theta_{\rm fermi}\}$, respectively. Numerical distinctions between the two strategies are discussed later toward the end of Section \ref{sec:numerics}.

\subsection{Training}\label{sec:training}

\subsubsection{Synthetic Data: Eigenvalue Sampling}
As described up to now, networks can be trained using real numbers inside the unit interval for energies instead of on matrices with normalized spectra. The method in which this scalar training data is sampled, however, has considerable impact on the accuracy of the optimized model. Regularly or randomly sampling according to some weighting does not affect the training, but random re-sampling between training iterations (analogous to using stochastic gradient descent instead of gradient descent) stalls performance, preventing convergence below a certain threshold of error. 

If we use a uniform distribution of eigenvalues as we perform the training, i.e.\ with a probability $p(x) \propto 1$, training acts to minimize the average error of the solution,
while sampling incorporating the derivative of the Fermi function, $p(x) \propto 1 + \left|f'\right|$, or proportional to the arc length, 
\begin{equation}
    p(x) \propto \sqrt{1 + (f')^2},
\end{equation}
acts, heuristically, to minimize the maximum error of the solution. At high values of $\beta$, it is necessary to incorporate a much higher rate of sampling (as satisfied by incorporating the derivative) around the Fermi level $\mu$, or else fine detail is missed and training error is not truly representative of the maximum error.

\subsubsection{Levenberg–Marquardt with Geodesic Acceleration}

The optimizer plays a critical role in the success of our overall approach to learning model coefficients: the problem is highly sensitive and not capable of \textit{exact} fits, but is nonetheless much closer to root-finding than to traditional gradient descent. For such problems, an efficient optimizer is nonlinear least squares, i.e. Levenberg-Marquardt.\cite{Transtrum2012} We initially trained our models in the previous section with the Julia package LsqFit.jl,\cite{lsqfitjl} before moving to the more recent NonlinearSolve.jl.\cite{NLSolvejl} 
Numerical improvements between the two packages as well as the introduction of simple geodesic acceleration constitute the difference between taking a million iterations to plateau in accuracy versus taking only a few hundred, while achieving an even greater accuracy.
With these improvements, training a model takes on the order of seconds to minutes, although this increases significantly with number of layers and sample points, both tied to higher $\beta$. However, as we see in the following section, all of our models are robustly transferable, so there is no need to retrain them once they are found. 

\section{Workflow}\label{sec:workflow}

Here we present a practical workflow for the algorithms developed in the previous sections and show how a single set of optimized Fermi-operator expansion coefficients can be used with expansions at various temperatures and fractional occupation. Our approach enables a model trained at a preselected $\beta_0$ and $\mu_0$ to be used within an entire region of normalized parameter space, which makes the methodology practical for simulations where $\beta$ or $\mu$ may change.

\subsubsection{Affine Rescaling to Avoid Retraining}
\label{ssec:rescaling}
We cannot train models for all possible $(\beta, \mu)$ pairs  and do not want to constantly re-fit a given model throughout the course of a dynamical simulation when a new temperature or chemical potential is needed. Fortunately, $\beta$ and $\mu$ appear as a linear transformation of the eigenvalue spectrum of $H$ in the Fermi-operator expansion. We can therefore rescale and shift $H$ to adjust the chemical potential and temperature. We leverage this by building a single model trained for parameters $(\beta_0, \mu_0)$, that is capable of representing an entire region of normalized $(\beta',\mu')$ parameter space with high accuracy. This region is called the region of validity. 

Given a true $H$, $\beta$ and $\mu$, the normalized parameters are
\begin{align}
    H'     & \equiv \frac{\varepsilon_\text{max} I - H}{\varepsilon_\text{max} - \varepsilon_\text{min}} \label{eq:H'}  \\
    \mu'   & \equiv \frac{\varepsilon_\text{max} - \mu}{\varepsilon_\text{max} - \varepsilon_\text{min}} \label{eq:mu'} \\
    \beta' & \equiv (\varepsilon_\text{max} - \varepsilon_\text{min}) \thinspace \beta \label{eq:beta'}
\end{align}
which ensure that
\begin{align} \label{eq:rescaling}
    &(H - \mu I) \thinspace \beta  \\  \nonumber
    & = \frac{H - \mu I}{\varepsilon_\text{max} - \varepsilon_\text{min}} \thinspace (\varepsilon_\text{max} - \varepsilon_\text{min}) \thinspace \beta \\ \nonumber
    & = \frac{H - \varepsilon_\text{max} I - \mu I + \varepsilon_\text{max} I}{\varepsilon_\text{max} - \varepsilon_\text{min}} \thinspace (\varepsilon_\text{max} - \varepsilon_\text{min}) \thinspace \beta \\  \nonumber
    & = \biggl( \frac{H - \varepsilon_\text{max} I}{\varepsilon_\text{max} - \varepsilon_\text{min}} - \frac{\mu - \varepsilon_\text{max}}{\varepsilon_\text{max} - \varepsilon_\text{min}} I \biggr) \thinspace (\varepsilon_\text{max} - \varepsilon_\text{min}) \thinspace \beta \\  \nonumber
    & = (\mu' I-H') \thinspace \beta'  \;.
\end{align}

In other words, $H, \beta, \mu$ and $H', \beta', \mu'$ generate the exact same Fermi-operator expansion, taking into account the spectrum flip used for the initial linear transformation of SP2. Similarly, if $\beta_0, \mu_0$ are the parameters we have trained our model to, we can determine an $H_0$ such that $H', \beta', \mu'$ and $H_0, \beta_0, \mu_0$ generate the same density matrix. Setting 
\begin{equation} \label{eq:H0}
 (\mu_0I-H_0)\beta_0 = (\mu'I-H')\beta' \;,
\end{equation}
leads to $H_0$ being given by
\begin{equation} \label{eq:H0}
    H_0 = \frac{\beta'}{\beta_0}(H'-\mu'I)+\mu_0 I\;.
\end{equation}
The only condition we need impose is that the spectrum of $H_0$ lie inside [0,1], a requirement of the models developed in Sec.~\ref{sec:algs}. Since this is already the case for $H'$ by construction, this condition on $H_0$ leads to the inequality constraints
\begin{equation}
    0 \le \frac{\beta'}{\beta_0}(0-\mu')+\mu_0 \quad \text{and} \quad  \frac{\beta'}{\beta_0}(1-\mu')+\mu_0 \leq 1 \;,
\end{equation}
or, equivalently,
\begin{equation} \label{eq:conditions}
    \frac{\mu_0}{\mu'} \beta_0 \geq \beta' \quad \text{and} \quad \frac{1 - \mu_0}{1 - \mu'} \beta_0 \geq \beta' \;.
\end{equation}

\begin{figure}[]
    \centering
    \includegraphics[width=\linewidth]{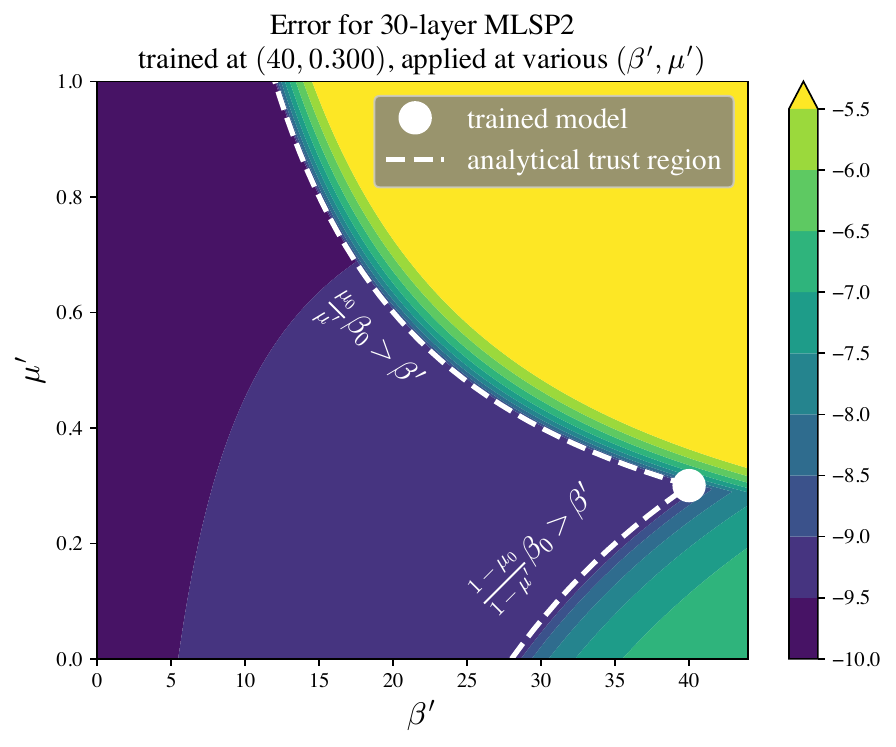}
    \includegraphics[width=\linewidth]{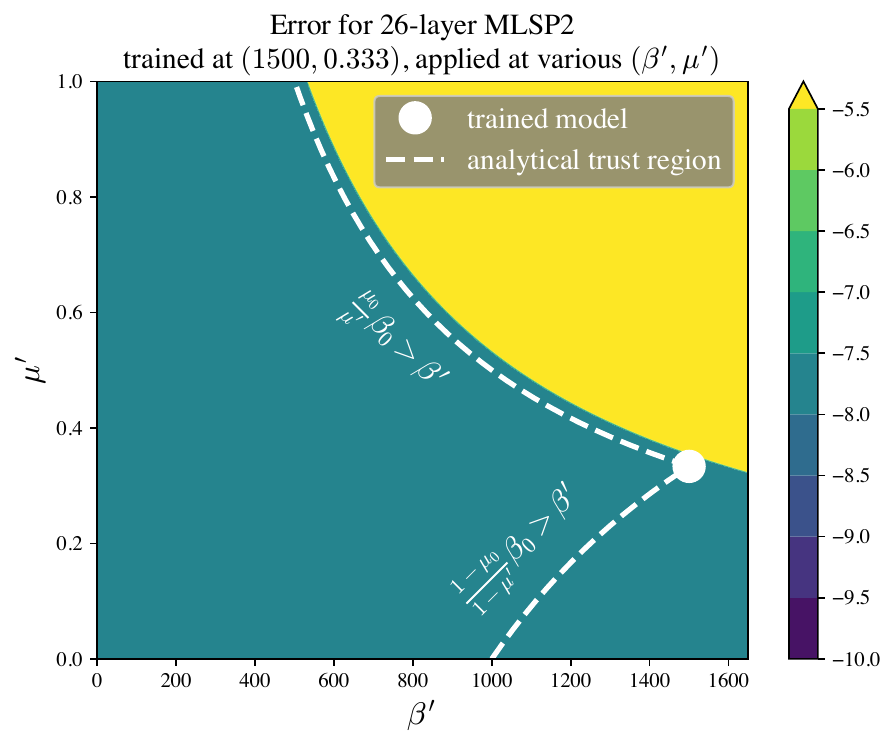}
    \caption{(top) Error of a model trained at $\beta_0 = 40, \;\mu_0 = 0.3$ when applied to a region of $\beta',\mu'$ normalized parameter space. The region of validity is the (blue) area in the plane with error (in the max norm) less than $10^{-9}$ to the left of dashed lines indicated by Eq.~(\ref{eq:conditions}). The negative exponent associated to each color, representing the model error, is shown on the right, with yellow regions having unbounded max error (at {\it least} $10^{-5.5}$). (bottom) Same plot but for $\beta_0=1500$ and $\mu_0=1/3$, with error allowed to go as high as $10^{-7.5}$ in the region of validity.}
    \label{fig:validity}
\end{figure}

These non-linear constraints define the region of validity of $(\beta', \mu')$ parameter space, for which the trained model with Hamiltonian $H_0$ and parameters $\beta_0, \mu_0$, has the same density matrix as $H'$, $\beta'$, $\mu'$, and thus of the original problem $H$, $\beta$ and $\mu$. The regions of validity for the example cases of $\beta_0 = 40, \mu_0 = 0.3$ and $\beta_0 = 1500, \mu_0 = 1/3$ are shown in Fig.~\ref{fig:validity}. 

The rescalings of temperature may require training $\beta_0$ for very high values in order to capture low electronic temperatures. The temperature scaling factor in Eq.~(\ref{eq:beta'}) will depend on the spectral range of the underlying Hamiltonian and should be taken into account when training a model at the point $(\beta_0,\mu_0)$. This idea is discussed further in Sec.~\ref{sec:region-by-layer} where we provide an upper bound on the accurately representable $\beta'$ values for a given $\beta_0$. In Alg.~\ref{alg:mlsp2} we display pseudocode which implements a calculation of the density matrix using the normalizations and rescalings discussed in this section along with the the entropy approximation from Eq.~(\ref{eq:entropy expansion}) and Eq.~(\ref{eq:entropy expansion-s}). A fully working python implementation of MLSP2 trained for the $(\beta_0,\mu_0)=(1500,1/3)$ model has also been included in Sec. A of the Appendix.

\begin{algorithm}
\caption{Using MLSP2 with normalizations to compute $D$ and $S$ at finite temperature} \label{alg:mlsp2}

Determine spectral bounds $\varepsilon_{\min}, \varepsilon_{\max}$ \; 

$\mu' = \frac{\varepsilon_{\max} -\mu}{\varepsilon_{\max} - \varepsilon_{\min}}$\;
$\beta' = (\varepsilon_{\max} - \varepsilon_{\min}) \beta$ \;

$H' = \frac{-1}{\varepsilon_{\max}-\varepsilon_{\min}} H + \frac{\varepsilon_{\max}}{\varepsilon_{\max} - \varepsilon_{\min}} I$ \; 
$ X = \frac{\beta'}{\beta_0}H' + \left( \mu_0 - \frac{\beta'}{\beta_0}\mu' \right) I $ \;
$ Y = \alpha X + (1-\alpha) \mu I $ \;

$A_D, A_S = 0, 0$ \;
\For{$1\leq i \leq n$}
{
    \tcp{$\theta$ refers to the model trained for the Fermi function, $\Theta$ to the model trained for entropy}
    $A_D = A_D + \theta_{i,4} \; X$ \;
    $X = \theta_{i,1} X^2 + \theta_{i,2} X + \theta_{i,3} I$ \;
    $A_S = A_S + \Theta_{i,4} \; Y$ \;
    $Y = \Theta_{i,1} Y^2 + \Theta_{i,2} Y + \Theta_{i,3} I$ \;
}
$D = A_D + X$\;
$Y = A_S + Y$\;
$S = (4 \ln 2) ( \Tr( Y^2 ) - \Tr(Y) )$ \;
\end{algorithm}

Just as the procedures discussed in this paper for training recursive quadratic expansions can be applied to any analytic function on a square matrix (so long as a good initial guess is provided), the procedure outlined in this section to reuse a model trained for a particular $(\beta,\mu)$ point can be generalized and used for any family of matrix functions where, like the Fermi function, the parameters only act in an affine manner on the input,
or where an affine transformation on the input matrix results in a similar transformation on the output,
\begin{equation}
    P(a X + b I) = Q(a, b) P(X) + R(a,b) I\;,
\end{equation}
where $Q, R$ can be any arbitrary functions, since $a$ and $b$ are scalars. Many potentially significant families of functions fall into this category, such as exponentials, $e^{aX+bI} = e^{aX}e^{bI} =e^b e^{aX}$, or, with $b,R=0$, square roots, $P(cX) = \sqrt{c} P(X)$, logarithms, $P(aX) = \ln(a)I + P(X)$, and orthogonalization,\cite{Muon} $P(cX)=P(X)$.

\subsubsection{Generalized SP2 Workflow}
\label{ssec:workflow}

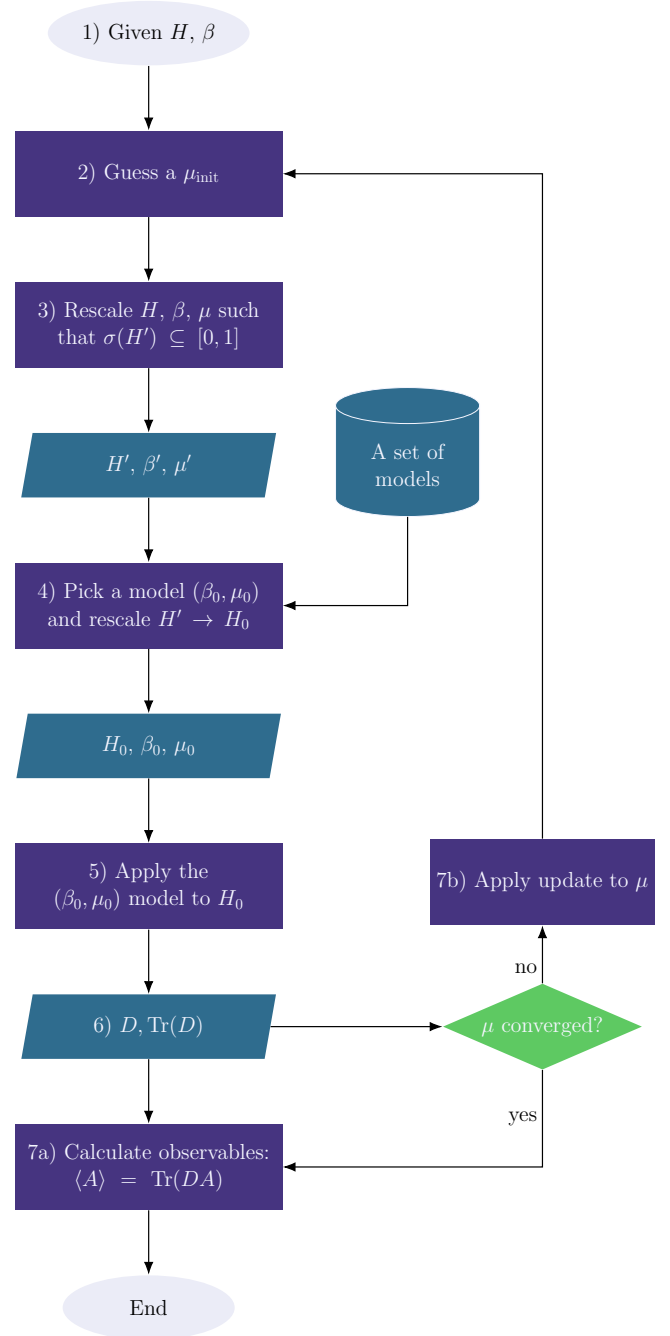
\begin{figure}[]
    \centering
    \resizebox{\linewidth}{!}{
    \begin{tikzpicture}[every node/.style={font=\Large}]
\node (start) [startstop] 
    {1) Given $H$, $\beta$};
\node (guess) [process, below=1.5cm of start, text width=6cm] 
    {2) Guess a $\mu_\text{init}$};
\node (rescale) [process, below=1.5cm of guess, text width=6cm]
    {3) Rescale $H$, $\beta$, $\mu$ such that $\sigma(H') \subseteq [0,1]$};
\node (model) [databox, below=1.5cm of rescale, text width=2.5cm]
    {$H'$, $\beta'$, $\mu'$};
\node (rescale2) [process, below=1.5cm of model, text width=6cm] 
    {4) Pick a model $(\beta_0,\mu_0)$ and rescale $H' \to H_0$};
\node (model2) [databox, below=1.5cm of rescale2, text width=2.5cm]
    {$H_0$, $\beta_0$, $\mu_0$};
\node (apply) [process, below=1.5cm of model2, text width=6cm] 
    {5) Apply the $(\beta_0, \mu_0)$ model to $H_0$};
\node (density) [databox, below=1.5cm of apply, text width=2.6cm]
    {6) $D, \Tr(D)$};
\node (property) [process, below=1.5cm of density, text width=6cm] 
    {7a) Calculate observables: \\ $\langle A \rangle = \Tr(DA)$};
\node (end) [startstop, below=1.5cm of property] 
    {End};
\node (decision) [decision, right=4cm of density] 
    {$\mu$ converged?};
\node (newton) [process, above=1.35cm of decision, text width=5cm] 
    {7b) Apply update to $\mu$};
\node (models) [database, right=1.5cm of model, text width=3cm] 
    {A set of models};

    \draw [arrow] (start) -- (guess);
    \draw [arrow] (guess.south) -- (rescale.north);
    \draw [arrow] (rescale.south) -- (model.north);
    \draw [arrow] (model.south) -- (rescale2.north);
    \draw [arrow] (rescale2.south) -- (model2.north);
    \draw [arrow] (model2.south) -- (apply.north);
    \draw [arrow] (apply.south) -- (density.north);
    \draw [arrow] (density.east) -- (decision.west);
    \draw [arrow] (decision.north) -- node[near start, left, text=black] {no} (newton.south);
    \draw [arrow] (newton.north) |- (guess.east);
    \draw [arrow] (density.south) -- (property.north);
    \draw [arrow] (property.south) -- (end.north);
    \draw [arrow] (models.south) |- (rescale2.east);
    \draw [arrow] (decision.south) |- node[near start, left, text=black] {yes} (property.east);
\end{tikzpicture}
    }
    \caption{This flowchart describes the process of applying the generalized MLSP2 algorithm to arbitrary Hamiltonians. Blocks demonstrate procedures, parallelograms contain intermediate data, the cylinder contains the database of available models, and the diamond represents the convergence criterion of some iterative method for the chemical potential.}
    \label{fig:flowchart}
\end{figure}

In this section we describe the step-by-step procedure for how to use any generalized MLSP2 method to compute the finite-temperature density matrix for a known occupation or chemical potential. Figure~\ref{fig:flowchart} condenses these steps into an easy-to-read flowchart. If $\mu$ is already known, there is no iterative update to $\mu$ and the outer loop in Fig.~\ref{fig:flowchart} is not needed.
\begin{enumerate}
    \item Obtain a Hamiltonian $H$ and electronic temperature $\beta$.
    \item  Furnish an initial guess for $\mu$ which is likely to be very close to the true one. 
    \item Estimate the spectral bounds of $H$ and rescale $(H, \beta, \mu) \rightarrow (H',\beta',\mu')$. 
    \item Pick a model $(\beta_0,\mu_0)$ that contains $(\beta',\mu')$ within its bounds according to \cref{eq:conditions}. $H'$ is then rescaled according to Eq.~(\ref{eq:H0}) to yield $H_0$.
    \item The model $(\beta_0,\mu_0)$ is applied to $H_0$ to compute the finite-temperature density matrix $D$.
    \item $D$ is obtained and if necessary, $\Tr(D)$ is checked against the desired occupation number $N_{\rm occ}$.
    \item[7a.] If these match sufficiently well, use $D$ to compute quantum observables of interest according to Eq.~(\ref{eq:observables}). 
    \item[7b.] If they do not, we update $\mu$ (e.g. through a Newton-Raphson step) and re-evaluate $D$ with a more precise $\mu$.
\end{enumerate}

If weights for multiple families of algorithms are stored, such as MLSP2, Skip-SP2, or Max-SP2, model selection (Step 4) can be done over each family, ultimately using one with the most favorable properties (such as minimum layer count subject to accuracy and memory constraints).

\subsubsection{Determining Chemical Potentials Using the Newton--Raphson Method}
\label{ssec:mu}

In electronic structure calculations, we often know the total number of electrons in the system, $N_\text{occ}$, {\it a priori} but not the chemical potential, $\mu$. Since the density matrix depends on $\mu$ it needs to be determined iteratively. To do so, we can solve the equation
\begin{equation}\label{eq:g}
   g(\mu) \equiv N_\text{occ} - \Tr(D) = 0\;
\end{equation}
using Newton-Raphson. \cite{ANiklasson08b} This requires a derivative of $g$, which in turn requires differentiating the density matrix with respect to $\mu$. The density matrix $D$ satisfies the logistic differential equation in $\mu$,
\begin{equation}\label{eq:drho}
   \frac{\partial D}{\partial \mu} = -\beta D (I - D)\;,
\end{equation}
since it is a logistic function of the Hamiltonian. If $D$ is approximated to reasonable precision, this expression can be used directly for $\partial D/ \partial \mu$, and is used in Ref.~\onlinecite{SMniszewski2019} once the initial chemical potential is determined. For MLSP2, $D$ is always approximated to a sufficient precision to enable use of Eq.~(\ref{eq:drho}), but previous work needed to evaluate the exact derivative for the first few Newton iterations, and computation via forward differentiation doubles the number of required matrix-matrix multiplications.

If Eq.~(\ref{eq:drho}) is used, $g'(\mu)$ can be evaluated with no additional matrix multiplications, since
\begin{equation}\begin{split} \label{eq:dg}
    g'(\mu) &= -\Tr \left( \frac{\partial D}{\partial \mu} \right) 
    = \beta \Tr(D) - \beta \Tr(D^2) \; ,
\end{split}\end{equation}
and 
\begin{align}
    \Tr(D) &= \sum_{i=1}^N D_{ii} \\
    \Tr(D^2) &= \sum_{i=1}^N \sum_{j=1}^N |D_{ij}|^2 \;. \label{eq:tr(D^2)}
\end{align}
This approach was used only as an approximation in Ref.~\onlinecite{SMniszewski2019}, but here it is exact up to machine precision. Thus the iterative correction to $\mu$ can be easily computed
\begin{equation}\label{eq:delta_mu}
   \Delta \mu = -\frac{g(\mu)}{g'(\mu)} = -\frac{N_\text{occ} - \Tr(D)}{\Tr\bigl(\beta D (I - D)\bigr)}\;.
\end{equation}
Because Newton-Raphson is quadratically convergent, with a sufficiently close guess, only one iteration of this procedure may be necessary. However, it should be noted that while this same procedure needs to be done for diagonalization-based methods, they incur the cost of diagonalization only once, at the decomposition of $H$, because \cref{eq:g,eq:delta_mu} can be evaluated and updated using purely the diagonal of the Hamiltonian. 

\section{Numerics} \label{sec:numerics}

\subsection{Tensor Cores and mixed-precision}
\label{sec:tensorcores}

For a symmetric matrix, $X$, represented in single-precision, the squaring operation $X \mapsto Y = X^2$ can be computed using mixed-precision arithmetic by decomposing $X$ into two half-precision summands, $X_0$ and $X_1$, and then expanding the product using FP16 matrix multiplications with FP32 accumulations (which is denoted as FP16/FP32), \cite{JFinkelstein2021}
\begin{equation} \begin{split} \label{eq:emulated matmul}
    X_0 &= \text{FP16}\;[X] \\
    X_1 &= \text{FP16}\;[X - \text{FP32}[X_0]] \\
    Y_0 &= X_0 X_0 \\
    Y_1 &= X_0 X_1 \\
    Y &\approx Y_0 +  \; Y_1 + \; Y_1^\top \; .
\end{split} \end{equation} 
We thus approximately obtain a square of an FP32 symmetric matrix using only two FP16/FP32 matrix-matrix multiplications. There is no equivalent trick for multiplying two non-symmetric $X_0$ and $X_1$ however, since in that case $Y_1^\top \neq X_1 X_0$, and three multiplications would be required.

Recently, Nvidia has integrated support for similar mixed-precision acceleration schemes into cuBLAS,\cite{cublas} which emulate higher precision multiplications using integer or low precision floating point data types on Tensor Cores based on the Ozaki scheme.\cite{HOotomo24} This emulation preserves both range and accuracy, but requires many more multiplications than the two needed in Eq.~(\ref{eq:emulated matmul}). Our approach leverages the inherent symmetry of our matrices, the fact that we require only a squaring operation and discard terms smaller than $2^{-22}$, the square of the FP16 unit round-off, which is both practically and rigorously justified. \cite{JFinkelstein2021,NHigham2021}  

An efficient approach to evaluating Eq.~(\ref{eq:mlsp2}) via Eq.~(\ref{eq:emulated matmul}) on such hardware can be given by Alg. \ref{alg:squaring}, where optimizations have been made for the sake of minimizing data transfer. 

\begin{algorithm} \label{alg:squaring}
    \caption{Mixed Precision MLSP2}
    $a,b,c,d = \theta_{i,1}, \; \theta_{i,2}, \; \theta_{i,3}, \; \theta_{i,4}$ \;
    $A = A + d \; X$ \;
    $X_0 = \text{float\_to\_half}(X)$ \;
    $X_1 = \text{float\_to\_half}(X-X_0)$ \;
    $X = aX_0X_0 + bX$ \;
    $X = aX_0X_1 + \tfrac{1}{2}X$ \;
    $X2 = X + X^\top$ \;
    $X = X2 + cI$ \tcp*[]{ $X_{i+1}=aX_i^2+bX_i+cI$ }
\end{algorithm}

Since Tensor Cores can give FP32 output (as a result of accumulating in FP32) there is no need to re-typecast to higher precision. The first line after generating $X_0$ and $X_1$ in Alg.~\ref{alg:squaring} gives $aY_0 + bX$, which is symmetric, the next line gives $aY_1 + \frac{1}{2}(aY_0 + bX)$, and the addition thereafter gives:
\begin{equation}
    \frac{1}{2}(aY_0 + bX) + \frac{1}{2}(aY_0^\top + bX^\top) + aY_1 + aY_1^\top
\end{equation} 
which, by symmetry again, reduces to:
\begin{equation}
a(Y_0 + Y_1 + Y_1^\top) + bX = aX^2 + bX\;.
\end{equation}
This removes the need for one matrix addition kernel compared to the naive implementation of Eq.~(\ref{eq:emulated matmul}). Further, $A = A+d\;X$ can be neglected when $d$ is less than some threshold, e.g. $10^{-8}$, which is the case for about half of layers, and training can be done with $c=0$ as the SP2 initialization has no constant term, replacing the last addition with a simple swap of pointers to $X$ and $X2$. 

All together, this reduction in data transfer improves efficiency at larger matrix sizes from $\sim73\%$ in the naive implementation to $\sim90\%$, resulting in a gain of $30$ Tflops on the RTX 6000 Ada.

\subsection{Layer Count} 
\label{sec:layer_count}
The learned weights of MLSP2 empirically tend to be similar to the original SP2 coefficients. This helps us to derive a reasonable estimate for the number of layers, $n_\ell$, required to train the MLSP2 model by looking at how large of a polynomial order is needed for the SP2-generated polynomial on [0,1] to approximate the true derivative of the Fermi function evaluated at $\mu'$, $f'(\mu')=\beta'/4$. Intuitively, we need a steep enough polynomial in order to be able to match derivatives at the chemical potential. 

Surprisingly, an estimate can be made as a natural consequence of the fact that the inverse of the golden ratio, $\varphi\approx 1.618$, is a fixed point of the composition of $f_0(x)=x^2$ with $f_1(x)=2x-x^2$. 

Setting $\phi = \varphi^{-1}$, it is simple to verify that $f_0(\phi) = \phi^2$, and that
\begin{align}
f_1(\phi^2) &= 2\phi^2 - \phi^4\\
&= \phi^{2}(2 - \phi^{2})\\
&= \phi^{2}((\underset{=\phi^{-1}}{\underbrace{\phi+1}}) - (\underset{=0}{\underbrace{\phi^2+\phi-1}}))= \phi \;,
\end{align}
so that $(f_1\circ f_0)(\phi)=\phi$. Note also that
\begin{align}
\begin{split}
    f_0'(\phi)   &= 2\phi \\
    f_1'(\phi^2) &= 2(1-\phi^2) = 2\phi \;.
\end{split}
\end{align}
If the initial linear transformation of SP2 in \cref{eq:sp2-mu} maps $\mu$ to $\phi$, an alternating sequence of $f_0$ and $f_1$ compositions 
\begin{align}
    \widetilde{f}_{n_\ell} \equiv f_{n_\ell \rm mod 2} \circ \cdots \circ f_1 \circ f_0 \;
\end{align}
approximates the step function (proof shown in the Appendix)
\begin{align}
    {\Theta}_j(x) = \begin{cases}
        0,     & x<\phi\\
        \phi^j , & x=\phi\\
        1,     & x>\phi
    \end{cases} \;,
\end{align}
on $[0,1]$ with $j=1$ if $n_\ell$ is odd, and $j=2$ if $n_\ell$ is even. We refer to this version of SP2, where we alternate between $f_0$ and $f_1$, as GoldenSP2. By the chain rule and induction, the derivative of $\widetilde{f}_{n_\ell}$ at $\phi$ is then
\begin{align}\label{eq:layer count no log}
\begin{split}
\widetilde{f}_{n_\ell}'(\phi) &= f_{n_\ell \rm mod 2}'(\phi^j) \frac{d}{dx}( f_{n_\ell-1 \rm mod 2} \circ \cdots \circ f_0)(\phi)\\
   &= (2\phi)^{n_\ell} \;,
\end{split}
\end{align}

Thus setting $\widetilde{f}_{n_\ell}'$ equal to the derivative of the Fermi function at $\mu' = \phi$ yields an estimate for $n_\ell$ of
\begin{equation} \label{eq:layer count}
    n_\ell = \frac{\ln(\beta'/4)}{\ln(2\varphi^{-1})} \approx 4.7 \ln \beta' - 6.5 \;.
\end{equation}
Empirically, we still observe this behavior on average for SP2 when $\mu' \neq \phi$ provided $\mu'$ is close to 0.5. 

\begin{figure}
   \centering
   \includegraphics[width=\linewidth]{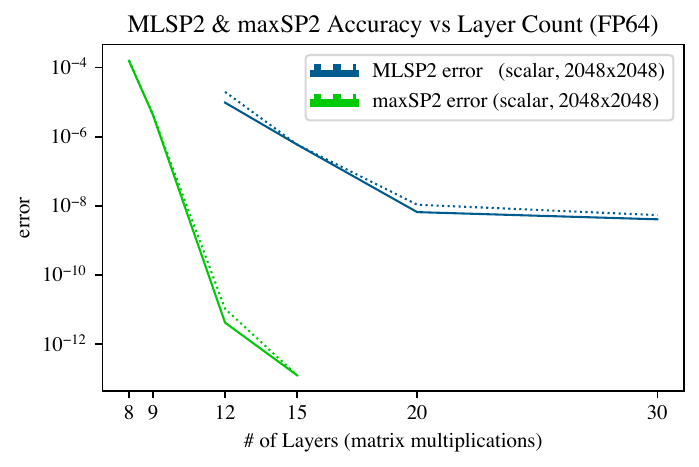}
   \caption{Model accuracy versus layer count at $\beta = 40,\; \mu = 0.3$, showing the error as a function of number of layers. In this case we are not limited by polynomial order, but rather expressibility. By Eq.~(\ref{eq:layer count}), $\beta=40$ only requires 11 layers to match derivatives.}
   \label{fig:accuracy vs layers}
\end{figure}

However, while this condition gives a minimum bound for layers needed to meet the slope requirement, it does not mean accuracy plateaus once this layer count is met. As Fig.~\ref{fig:accuracy vs layers} shows, beyond some baseline level, accuracy can further increase with expressibility or number of parameters, either by architecture change such as Max-SP2 or additional layers for MLSP2. This is unnecessary, however, in the regime where numerical precision is the dominant cause of error, as opposed to error in the underlying model (see Fig.~\ref{fig:hardware_accuracies}).

\subsection{Region of Validity By Layer Count}\label{sec:region-by-layer}

If $0<\mu'<1/2$, the inequalities of Eq.~(\ref{eq:conditions}) can be used to bound $\beta'$ by
\begin{equation} \label{eq:betamax}
    2\mu_0 \beta_0 \geq \beta' \quad \text{and} \quad (1 - \mu_0) \beta_0 \geq \beta' \;.
\end{equation}
Since both inequalities need to be satisfied, the bounds are maximal when $\mu_0=1/3$, leading to 
\begin{equation}
    \beta'\leq \tfrac{2}{3} \beta_0 \;.
\end{equation}
Making use of the symmetry of $f$ 
\begin{equation} \label{eq:flip}
    f(x'; \beta', \mu') = 1 - f(1-x'; \beta', 1-\mu') \;,
\end{equation}
on the unit interval extends the bound to all $0<\mu'<1$. By also using Eq.~(\ref{eq:layer count no log}), we are able to deduce that for any desired layer count $n_\ell$, a model can be trained at
\begin{equation} \label{eq:train-model-at}
\beta_0 = 4(2\phi)^{n_\ell},\quad  \mu_0 = \frac{1}{3}\;, 
\end{equation}
and can then be used with any $0<\mu'<1$ and $0<\beta' \leq \tfrac{2}{3} \beta_0$ while still maintaining the same accuracy. This makes it possible to train a single set of coefficients (or, weights) that is usable for a wide range of temperatures and chemical potentials. For example, one could train a 10 layer model up to $\beta_0 = 33.3$, and thus use only 10 layers for any Fermi evaluation up to $\beta' = 22.2$. The estimate given in Eq.~(\ref{eq:train-model-at}) indeed suggests that higher values of $\beta_0$ do require more layers, an observation we also make empirically. At sufficiently large $\beta_0$, however, it may not be possible to obtain a set of weights yielding an accurate MLSP2 model, necessitating a more expressive alternative architecture such as Max-SP2.

\begin{figure}
   \centering
   \includegraphics[width=\linewidth]{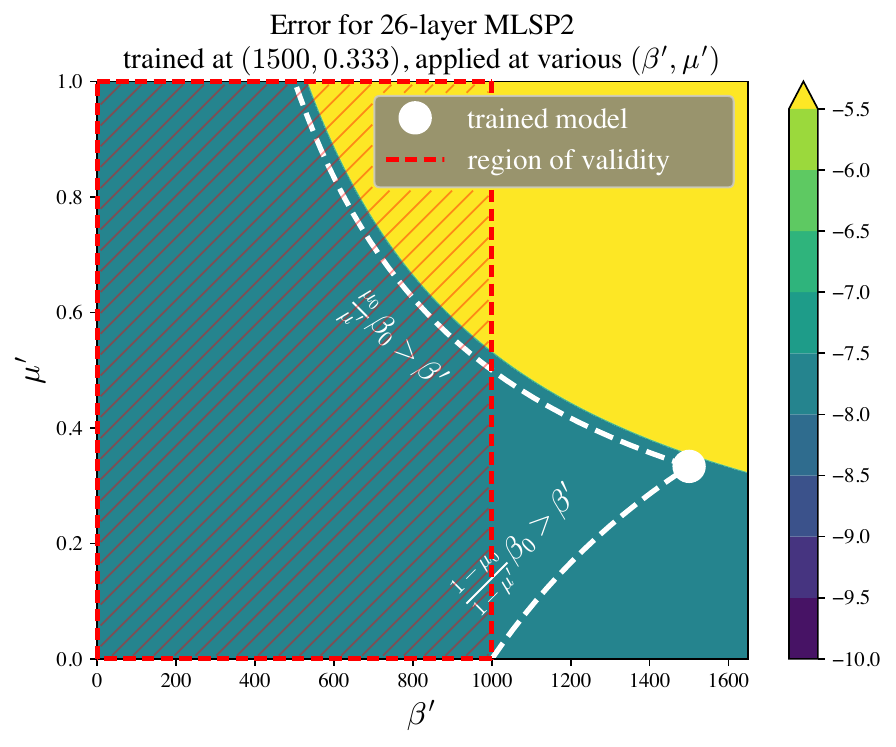}
   \caption{Same figure as the one shown in Fig.~\ref{fig:validity} for $\beta_0 = 1500$ and $\mu_0 = 1/3$ now with the region of validity depicted inside the red rectangle. The accuracy at any $\beta', \mu'$ inside the red region of validity is guaranteed to be the same or better as the accuracy of the model trained at $\beta_0=1500$ and $\mu_0=1/3$.}
   \label{fig:region_of_validity}
\end{figure}

Going back to the $\beta_0 = 1500$ and $\mu_0 = 1/3$ model from Sec.~\ref{ssec:rescaling}, we now see that these parameter choices enable model use for all $0 < \beta' \leq 1000$ and all valid $\mu'$, i.e. $0<\mu'<1$. This region of validity for $\beta_0=1500$ and $\mu_0=1/3$ is shown inside the red rectangle in Fig.~\ref{fig:region_of_validity}. By Eq.~(\ref{eq:conditions}) and the fact that $\beta = 1/k_B T$, this model can therefore handle a temperature as low as  
\begin{align} \label{eq:temp limit}
    T \ge \frac{(\varepsilon_{\rm max}-\varepsilon_{\rm min})}{k_B \times 1000} = 11.6(\varepsilon_{\rm max}-\varepsilon_{\rm min}) \; {\rm K}\;.
\end{align}

Thus, if the spectral width of $H$ is on the order of $1$ eV, the model can very accurately capture temperatures as low as 11.6 K. However if the spread in energies is on the order of 100 eV, the model cannot reliably go below 1160 K, and capturing lower temperatures would require a new model be trained at a higher $\beta_0$. For some atomic-orbital basis sets or semi-empirical approaches (such as Density Functional Tight-Binding\cite{MElstner98}), the energy difference $\varepsilon_{\rm max}-\varepsilon_{\rm min}$ is typically on the order of 10 to 100 eV. Our model, trained with $\beta_0 = 1500$ and $\mu_0=1/3$, should therefore be applicable to a broad range of practical, finite-temperature simulations employing these methods.

\subsection{Comparison to Diagonalization}
In many practical settings, diagonalization is commonly used to construct the finite-temperature density matrix, making it a meaningful point of comparison. For example, in smaller-scale, conventional calculations using atomic-orbital basis sets, the density matrix can often be obtained by directly diagonalizing the full Hamiltonian. Larger-scale approaches, such as graph-based linear-scaling electronic structure methods,\cite{ANiklasson16,MKulichenko25,CLi26}  construct a sparse density matrix by independently evaluating the Fermi operator on many overlapping, moderately-sized dense principal submatrices of the Hamiltonian defined by core–halo subgraphs, and then assembling their nonoverlapping core contributions in parallel. Other large-scale approaches using systematically convergent basis sets, such as plane waves, real-space grids, or high-order finite elements, typically rely on iterative diagonalization techniques,\cite{PMotamarri20,GKresse99} which project the full Hamiltonian down onto a small subspace. Although the full discretized Hamiltonian may span an extremely broad energy range, the projected (or reduced) Hamiltonian can have a much narrower spectrum, potentially on the order of $10^2$ eV or less, making these reduced Hamiltonians suitable for the recursive Fermi-operator expansions schemes presented in this paper.\cite{SKhadatkar23} Our finite-temperature expansion schemes should therefore provide an attractive alternative to diagonalization for all of these these various use cases.

Our numerical comparisons focus on the MLSP2 method. Using random, symmetric Hamiltonian matrices we compute the single-particle density matrix using both diagonalization and MLSP2, and then compare wall clock times. Our experiments were run on an Nvidia RTX 6000 Ada GPU using both single and double precision for diagonalization and our mixed precision Tensor Core approach from Sec.~\ref{sec:tensorcores} for MLSP2. The MLSP2 model was trained at parameters $\beta_0 = 1500$ and $\mu_0=1/3$ and the number of layers was calculated using Eq.~(\ref{eq:layer count}) and resulted in 28 layers. With as few as 26 layers, the overall model error was less than the single precision unit-roundoff of $2^{-24}$.

\begin{figure}[H]
    \centering
    \includegraphics[width=\linewidth]{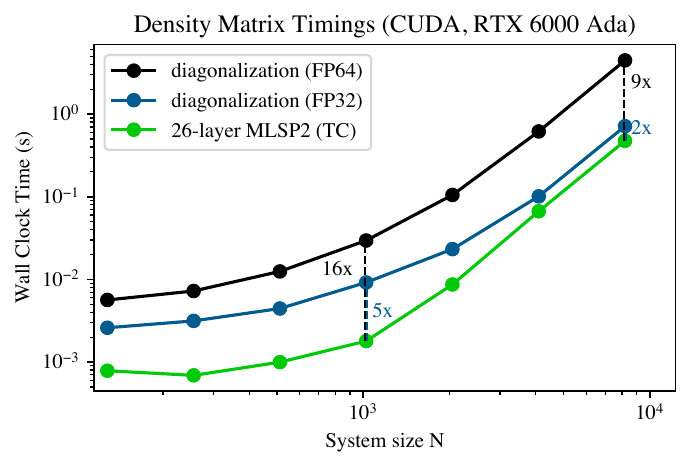}
    \caption{Wall clock time for a 26-layer MLSP2 evaluation (up to $\beta' \leq 1000$) using Nvidia Tensor Cores compared to the single and double precision cuSolver divide-and-conquer diagonalization, \texttt{cusolverDn<T>syevd}, on an Nvidia RTX 6000 Ada. }
    \label{fig:benchmarks}
\end{figure}

The results seen in Fig.~\ref{fig:benchmarks} confirm that MLSP2 evaluated using Nvidia Tensor Cores offer significant speedups over diagonalization-based density matrix construction. Diagonalization was evaluated using cuSOLVER divide-and-conquer routine for symmetric matrices, \texttt{cusolverDn<T>syevd}, which does not use Tensor Cores. $\texttt{<T>}$ can be either $\texttt{D}$ or $\texttt{S}$ for double or single precision, respectively. At intermediate matrix sizes, the figure shows a 16x speed for MLSP2 over a double precision diagonalization-based density matrix construction, and a 9x speed up for larger matrix sizes. 

Comparing to single precision diagonalization instead results in smaller speedups of 5x and 2x. However, divide-and-conquer diagonalization can be unreliable in single precision, sometimes producing much larger errors than MLSP2 when compared to double precision diagonalization, or diverging entirely, with no flag or indication of failure. Thus it is reasonable to consider double precision diagonalization as the more meaningful comparison.

The MLSP2 algorithm, like the other models we have developed, include both quadratic and cubic scaling components. Although the dominant computational kernel consists of dense matrix–matrix multiplications, each iteration of MLSP2 also requires substantial matrix preparation for low-precision arithmetic (including type conversion and data movement) as well as Level-2 BLAS-like operations such as matrix additions. For matrix sizes $N \leq 1024$ in Fig.~\ref{fig:benchmarks}, wall-clock time is dominated by these quadratic scaling operations, instead of the higher, cubic-scaling complexity Tensor Core multiplications. 

We can also estimate MLSP2 performance relative to diagonalization at large $N$ by fitting the MLSP2 data for $N>1024$ and the FP64 diagonalization results to a cubic scaling law. The resulting cubic prefactor for MLSP2 is approximately an order of magnitude smaller than that of FP64 diagonalization, suggesting that this 26-layer MLSP2 method will remain roughly an order of magnitude faster in the limit of large $N$.

The performance gains described are also hardware dependent. An Nvidia RTX 6000 Ada has a peak Tensor Core performance of about 180 Tflops for FP16 data, \cite{rtx6000} whereas regular, single-precision (non-Tensor-Core) performance on this device is about 90 Tflops. This results in a Tensor Core to single-precision performance ratio of about 2:1. On more high-performance computing oriented devices, such as an Nvidia H100, this ratio is closer to 15:1. \cite{h100} We thus expect a much greater performance difference in favor of MLSP2 on these kinds of devices. 

\begin{figure}[]
    \centering
    \includegraphics[width=\linewidth]{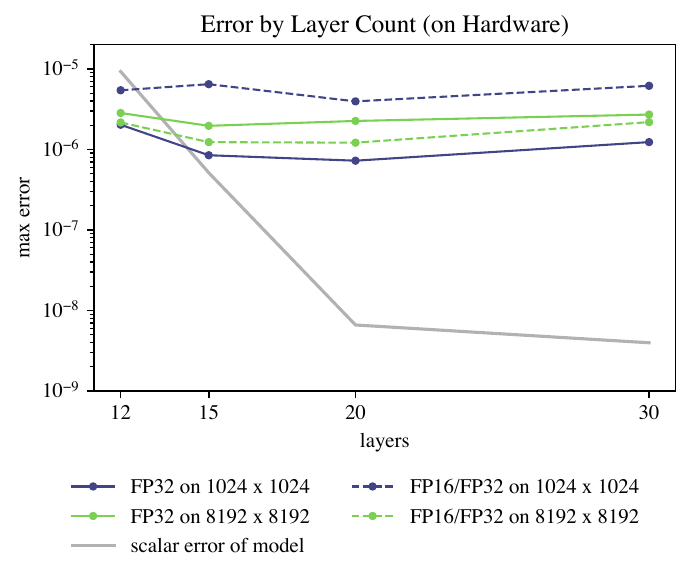}
    \caption{Model error compared to our accelerated mixed precision algorithm on Tensor Cores (dashed lines) and standard single precision (solid lines). We hit the maximum possible FP32 accuracy once the layer count satisfies Eq.~(\ref{eq:layer count}). The model was trained at $\beta_0=40$, $\mu_0 = 0.3$.}
    \label{fig:hardware_accuracies}
\end{figure}

Our speed advantage comes at an accuracy tradeoff. When competing with high-performance diagonalization routines, accuracy is determined not only from the model accuracy, as seen in Fig. \ref{fig:hardware_accuracies}, but also from the underlying precision used in the matrix multiplication implementation. In the case of using single precision or the Tensor Core implementation from Sec.~\ref{sec:tensorcores} in our models, we are limited to accuracies between $10^{-5} \text{ and } 10^{-6}$ (measured in the 2-norm), regardless of how accurate the underlying model is.

Practically, some of the advantage is also dampened by the need to iteratively determine $\mu$ in the case when it is not known a priori, as it is defined implicitly through the sum of the fractional occupation numbers equaling the occupation number.  Computing the density matrix with diagonalization, however, only requires a single diagonalization. This is because the energies, $\varepsilon_i$, are then known and we only need to solve the one-dimensional equation $\sum_i f_i =\Tr(D)=N_{\rm occ}$ 
to compute the unknown $\mu$. Nevertheless, as long as the number of iterations for determining $\mu$ is small, typically 1 or 2 in many practical simulations, we expect to maintain a sizable speed advantage over diagonalization.

\subsection{The Approximate Entropy Calculation}

We now bring our focus back to the approximate entropy function, which we denote by $\widetilde{s}$ below, and discuss the numerical accuracy implications. Table~\ref{tab:entropy} shows a numerical comparison for various model approximations of the entropy for a test, normalized Hamiltonian matrix, $H'$, with a uniformly distributed set of 2048 energy eigenvalues on the unit interval. The second column shows the maximum entropy error that could arise for an individual eigenstate, i.e. $\max|s(x) - \widetilde s(x)|$ for all normalized energies $x\in[0,1]$. The third through fifth columns show the relative error in entropy, i.e. 
\begin{equation}
    \frac{|\Tr [s(H')] - \Tr [\widetilde{s}(H')]|}{\Tr [s(H')]} = \frac{|S - \widetilde{S}|}{S}\;,
\end{equation}
using different floating point precisions in the expansions. The ``mixed" column here again refers to our FP16/FP32 approach described earlier in Sec. \ref{sec:tensorcores} and the FP32 and FP64 columns refer to case where all matrices are stored and operated on in FP32 or FP64, respectively. The last column describes the number of matrix multiplications needed to estimate $S$ if $D$ is known, accounting for the fact that $\Tr(D^2)$ does not require a matrix multiplication by Eq.(\ref{eq:tr(D^2)}). Here, $n_\ell=26$ and is the number of layers used in MLSP2. 

In the last two rows of Table~\ref{tab:entropy}, we again use the notation $\theta_\text{fermi}$ and $\theta_\text{entropy}$ to refer to different variations of the entropy model described in Eq.~(\ref{eq:entropy expansion}) and Eq.~(\ref{eq:entropy expansion-s}). The first variation has model coefficients $\{\theta_\text{fermi}\}$, and simply uses the predetermined MLSP2 weights for the Fermi function approximation (optimized for the same $\beta$ and $\mu$) together with the $\alpha$-rescaled ansatz based on Eq.\ (\ref{eq:approx-s}) with $m = 1$ and $\alpha = 0.842704$. No further training is performed. The second variation, trains a new set of model coefficients by using the pretrained MLSP2 weights only as an initial guess. The weights for this model are denoted by $\{\theta_\text{entropy}\}$.

From the table, these models show a clear computational cost versus accuracy tradeoff, including how the accuracy depends on the arithmetic floating point precision used in the matrix calculations. The mixed-precision approach, rows 4 and 5 of Table~\ref{tab:entropy}, only achieves a relative accuracy on the order of $10^{-4}$, which is still however, orders of magnitude better than a high-precision evaluation using Eq.~(\ref{eq:approx-s}) with different choices of $m$ (rows 1 through 3). Interestingly, the table shows an obvious preference for $\theta_{\rm entropy}$ over $\theta_{\rm fermi}$. The extra training performed to generate the weights $\{\theta_{\rm entropy}\}$ results in superior accuracy in all precisions compared to $\theta_{\rm fermi}$. For maximum accuracy of the custom learned entropy model however, the multiplications for $\theta_{\rm entropy}$ require FP64 precision.

\renewcommand{\arraystretch}{1.5}
\begin{table}[]
   \centering
   \begin{tabular}{c|c|c|c|c|c}
                 & Model  & \multicolumn{3}{c|}{\ Rel. err. of $\widetilde{S}$ \ } & Mults. \\
         Method  & max err. & \ mixed \ & \ FP32 \ & \ FP64 \ & needed \\
        \hline
        Eq.(\ref{eq:approx-s}), $m=1$        & 7.55e-2 & 0.157 & 0.157 & 0.157 & 0 \\
        Eq.(\ref{eq:approx-s}), $m=2$        & 3.71e-2 & 0.080 & 0.079 & 0.079 & 1 \\
        Eq.(\ref{eq:approx-s}), $m=4$        & 1.82e-2 & 0.040 & 0.039 & 0.039 & 2 \\
        Eq.(\ref{eq:entropy expansion}), $\theta_\text{fermi}$  & 1.24e-3 & 3.4e-4 & 6.4e-5 & 6.8e-5 & $n_\ell$ \\
        Eq.~(\ref{eq:entropy expansion}), $\theta_\text{entropy}$ & 6.73e-8 & 2.4e-4 & 4.4e-6 & 1.4e-8 & $n_\ell$ \\
   \end{tabular}
   
   \caption{Error comparison of entropy approximation methods for $\beta=40, \mu = 1/3$ with energies $x$ in the unit interval. The last column refers to how many matrix multiplications are needed to compute the trace of the entropy, $\Tr[\widetilde{S}]$, assuming that the density matrix $D$ has already been calculated via the method indicated by the row. }
   \label{tab:entropy}
\end{table}

\section{Conclusion}
We developed several finite-temperature recursive Fermi-operator expansion schemes based on various neural network formulations of the second-order spectral projection (SP2) method, thereby generalizing SP2 to finite electronic temperatures. The expansion coefficients are determined using machine learning models that offer different levels of expressivity, numerical accuracy, and computational cost. Furthermore, once a model is trained for a sufficiently low temperature, the model becomes robustly transferrable and can be used in a broad region of normalized parameter space. This makes our models a convenient choice for dynamical finite temperature simulations, such as quantum molecular dynamics, where the chemical potential and eigenvalue distribution may fluctuate between time steps. Our approach avoids the need for diagonalization and relies solely on highly optimized matrix-matrix multiplication kernels. In addition to the density matrix calculation, we also showed how a corresponding diagonalization-free approach could be used to provide accurate predictions of the electronic entropy. We demonstrate how these schemes can be implemented and applied to construct density matrices on both GPUs and dense matrix multiply units (e.g. Nvidia Tensor Cores) with mixed precision algorithms, showcasing large speedups relative to vendor optimized diagonalization routines. A high performance CUDA implementation of MLSP2 is contained in the \texttt{src/sedacs/gpu/nvda} directory within the github repository: \url{https://github.com/lanl/sedacs}.

\section*{Supplementary Material}
A working Julia code for training an MLSP2 model has been included as supplementary information. This code can be used to train an MLSP2 model for different values of $\beta_0$. 

\section*{Acknowledgments}
This work is supported by the U.S. Department of Energy, Office of Basic Energy Sciences (FWP LANLE8AN) and by the Laboratory-Directed Research and Development Program (20240006DR) at Los Alamos National Laboratory (LANL). LANL is operated by Triad National Security, LLC, for the National Nuclear Security Administration of the U.S. Department of Energy Contract No. 892333218NCA000001. This document has LA-UR number LA-UR-26-22648.

\section*{Author declarations}
The authors have no conflict of interest to disclose.

\section*{Data availability}
All codes for MLSP2, including neural network weights and the mechanism for their generation, are included in the text, appendix, or supporting information. Julia training code for Max-SP2 can be made available upon request.
 
\section*{References}
\bibliography{ref}

\vfill
\pagebreak

\appendix

\newcommand{\commentoutA}[1]{}
\renewcommand{\r}{\mathbf{r}}
\newcommand{\V}{\mathbf{V}}
\newcommand{\F}{\mathbf{F}}
\newcommand{\W}{\mathbf{W}}
\newcommand{\X}{\mathbf{X}}
\newcommand{\B}{\mathbf{B}}
\newcommand{\TODO}[1]{\textcolor{red}{#1}}
\makeatletter

\definecolor{codegreen}{rgb}{0,0.6,0}
\definecolor{codegray}{rgb}{0.5,0.5,0.5}
\definecolor{codepurple}{rgb}{0.58,0,0.82}
\definecolor{backcolour}{rgb}{0.95,0.95,0.92}

\lstdefinestyle{mystyle}{
    backgroundcolor=\color{backcolour},   
    commentstyle=\color{codegreen},
    keywordstyle=\color{magenta},
    numberstyle=\tiny\color{codegray},
    stringstyle=\color{codepurple},
    basicstyle=\ttfamily\footnotesize,
    breakatwhitespace=true,         
    keepspaces=true,                 
    numbers=none,
    showspaces=false,                
    showstringspaces=false,
    columns=fixed,
    tabsize=4,
    basewidth=0.5em,
}

\lstset{style=mystyle}

\begin{widetext}
\section{MLSP2 python implementation}\label{sec:python-code}
Here we present a working Python implementation of MLSP2 complete with trained model coefficients for $\beta_0=1500$ and $\mu_0=1/3$. This code uses double precision and needs to be properly adapted to take advantage of the mixed-precision formulation presented in the main text.
\begin{lstlisting}[language=Python,mathescape=true]
import numpy as np

# Model Weights for MLSP2 defined at $\color{codegreen} \beta_0 = 1500, \; \mu_0 = 1/3$

#          $\color{codegreen}\theta_{i,1}$                       $\color{codegreen}\theta_{i,2}$                     $\color{codegreen}\theta_{i,3}$                     $\color{codegreen}\theta_{i,4}$
weights = [
    [ -1.353996666080607e+00, +2.437540893737664e+00, -2.038466804171965e-01, +1.291156147536192e-11 ],
    [ +1.608677030263155e+00, -2.987544634415441e-01, -6.746741346599809e-02, -7.006084032292678e-12 ],
    [ -2.351636735892118e+00, +2.613813908880740e+00, -2.081115324436639e-03, -6.572840812953572e-13 ],
    [ +1.810031371507046e+00, -3.733036049410451e-02, -5.442179841399564e-02, +4.318503541775794e-11 ],
    [ -2.732204264661287e+00, +2.825886342002125e+00, +8.167975559885774e-03, -1.662456055798004e-11 ],
    [ -1.873987421378235e+00, +2.138798318408699e+00, +3.573114610928145e-02, +4.542623791264443e-12 ],
    [ +1.402261079424850e+00, +4.890736627204146e-02, -4.237220339751432e-02, +2.424648524635838e-11 ],
    [ +1.713089400763156e+00, -1.034037621506357e-01, -2.438094942357169e-02, +7.873381643056254e-11 ],
    [ -3.252595736449542e+00, +2.628649626717414e+00, +3.165347918592593e-02, -6.312379678783267e-11 ],
    [ -1.878046620274372e+00, +2.074880349608742e+00, +3.449163838903798e-02, -6.649574392039178e-11 ],
    [ +1.310464008049612e+00, +5.496499701340592e-02, +3.310087064751345e-03, +2.825980776677432e-10 ],
    [ +1.563608995897714e+00, -8.606990671650539e-03, -1.301065468049911e-02, -1.424083924407828e-07 ],
    [ -3.038983738437103e+00, +2.477686862799397e+00, +2.636871201950337e-02, -1.330100571490071e-07 ],
    [ -1.699377329573085e+00, +2.076565284757246e+00, +3.442469802192470e-02, +4.896446563691430e-03 ],
    [ +1.256940670391469e+00, +5.777573303514125e-02, +4.902596975631270e-02, -2.302054912581413e-03 ],
    [ -1.390281989291434e+00, +1.966243706278270e+00, +1.910503489826946e-02, +1.606259300834474e-02 ],
    [ +1.172415492052006e+00, +2.982302801604828e-02, +4.530045685625948e-02, -1.539021333689731e-02 ],
    [ +1.351448045975672e+00, +5.773064380022182e-02, +3.753306163777001e-02, -1.276795931125504e-01 ],
    [ -1.212734900653146e+00, +1.861660915086903e+00, +1.242573869144004e-02, -1.100276149057915e-01 ],
    [ +8.610663421144649e-01, +1.048747203546148e-02, +4.701135000272372e-02, -4.946843804326570e-02 ],
    [ -1.369173339062334e+00, +1.699077704335121e+00, -7.732969926664974e-03, -1.761228530611925e-01 ],
    [ -1.452079661340262e+00, +1.993171071701545e+00, -1.661115887219701e-02, -5.688902388963853e-01 ],
    [ +1.347262922911275e+00, +2.198074427424086e-02, +1.995615256168758e-01, +5.964487309001580e-01 ],
    [ -1.114238353358810e+00, +1.896937623313116e+00, -7.280733281872145e-02, +1.530915295291056e+00 ],
    [ +9.411010738180853e-01, -1.554719335561439e-01, +6.584253835393120e-02, -2.059339861429303e-01 ],
    [ +8.321693583250559e-01, +5.698259888759222e-01, -4.234806053199446e-01, +5.698259888760550e-01 ]
]

beta0, mu0 = 1500, 1/3

# Gershgorin circle theorem providing minimum and maximum bounds for a real spectrum
def Gershgorin(A):

    # extract diagonal entries of matrix (NOT diagonalization)
    D = np.diag(A)

    # radii of Gershgorin circles for each row
    R = np.sum(np.abs(A),axis=1) - np.abs(D)  

    # return minimum and maximum of all bounds of all circles
    return np.min(D-R), np.max(D+R)

# defined for $\color{codegreen}(e_{\max} - e_{\min}) \times \beta < 1000, \; e_{\min} < \mu < e_{\max}$
def mlsp2(H, beta, mu):

    # estimate bounds on spectrum of true H
    emin, emax = Gershgorin(H)
    
    # Identity matrix
    I = np.eye(H.shape[0])

    # primed variables given by Eqs. $\color{codegreen} \text{\ref{eq:H'} - \ref{eq:beta'}}$
    H_prime    = (emax * I - H)/(emax - emin)
    mu_prime   = (emax - mu)/(emax - emin)
    beta_prime = (emax - emin) * beta

    # flip given by Eq. $\color{codegreen} \text{\ref{eq:flip}}$ if mu' > 0.5
    mu_switch = mu_prime > 0.5
    if mu_switch:
        H_prime  = I - H_prime
        mu_prime = 1 - mu_prime

    # condition for validity by Eq. $\color{codegreen} \text{\ref{eq:conditions}}$
    beta_max = min( mu0/mu_prime, (1 - mu0)/(1-mu_prime) ) * beta0
    if beta_prime > beta_max:
        # Warn of temperture limit given by Eq. $\color{codegreen} \text{\ref{eq:temp limit}}$
        print("Warning: Temperature / Spectral Width of H is too low!") 
        print(f"limit = {1/(8.617e-5 * min(2*mu0,1-mu0) * beta0):.1f}K/eV.")
        beta_prime = beta_max
    
    # H0 given by Eq. $\color{codegreen} \text{\ref{eq:H0}}$
    X = (H_prime - mu_prime * I)*(beta_prime/beta0) + mu0 * I
    A = np.zeros_like(X)

    # MLSP2 as given by Eq. $\color{codegreen} \text{\ref{eq:mlsp2}}$
    for a,b,c,d in weights:
        A += d * X
        X = a * np.matmul(X,X) + b * X + c * I
        # for faster routines in cuBLAS see Sec. $\color{codegreen} \text{\ref{sec:numerics}}$ Numerics
    
    # corresponding flip given by Eq. $\color{codegreen} \text{\ref{eq:flip}}$
    return I - (A + X) if mu_switch else (A + X)


# test cases
if __name__ == "__main__":

    for i in range(10):

        # seed random number generation for reproducibility and fix matrix size
        rng = np.random.default_rng(seed=i)
        N = 256

        # generate random matrix of size (N,N), make it symmetric and more strongly diagonal
        H = rng.uniform(0, 1, (N,N))                            
        H = H + H.T
        $\text{H += np.diag(rng.uniform(0, N ** 0.5, (N)))}$

        # estimate spectral bounds
        emin, emax = Gershgorin(H)

        # choose beta and mu such that $\beta'$ and $\mu'$ fall inside region of validity
        beta = rng.uniform(1, (2/3 * beta0)/(emax - emin))
        $\text{mu   = (emax + emin)/2} $

        # construct density matrix with MLSP2
        D_mlsp2 = mlsp2(H, beta, mu)

        # compare to diagonalization
        evals, evecs = np.linalg.eigh(H)                     
        occupation = 1 / (1 + np.exp(beta * (evals - mu)))
        D_diag = (evecs * occupation ) @ evecs.T

        # print matrix 2-norm of error (maximal eigenvalue difference)
        print( np.linalg.norm(D_diag - D_mlsp2, ord=2) )
\end{lstlisting}
\end{widetext}

\section{Analysis of Spectral Refinement Networks}
\label{appendix:Analysis of Spectral Refinement Networks}

The application of analytic functions to matrices, such as the application of the Fermi operator to a discretized Hamiltonian matrix, is trivially equivariant. By this we mean there is a group of similarity transformations, and indeed any basis transformations, which commute with the function application
\begin{equation}\label{eq:basis transformation}
    f(SHS^{-1}) = Sf(H)S^{-1} \; .
\end{equation}
Such equivariance of electronic structure calculations under basis changes is natural and taken for granted. However, this may be of interest considering recent work in
Equivariant Neural Networks,\cite{Batzner2022} defined as neural networks whose application commutes with a fixed symmetry group $G$. By that definition, the network $f: X \rightarrow Y$ going from a vector space $X$ to a vector space $Y$ is $G-$equivariant if it satisfies\cite{Allen2026}
\begin{equation} \label{eq:ENNs}
    f(D_X[g] x) = D_Y[g] f(x), \qquad \forall x \in X, \forall g \in G \; .
\end{equation}
That is, for every element $g$ of group $G$, the function application commutes with the representations $D_V[g]$ of $g$ in each vector space $V\in\{X,Y\}$. Eq.~(\ref{eq:basis transformation}) can thus be considered as a form of equivariance in the sense of Eq.~(\ref{eq:ENNs}) over the real vector space of Hermitian matrices, 
\begin{equation}
    \vec{f}((S^{-T} \otimes S)\text{vec}(H)) = (S^{-T} \otimes S) \vec{f}(\text{vec}(H))\;.
\end{equation}
Since $S$ can be any basis transformation, this forms an equivariant neural network under the general linear group $G=GL(N)$. For those so inclined, this makes the representation $D_V$ the adjoint representation of $G$, i.e. the fundamental $\otimes$ antifundamental representation.

Our method further deviates from traditional neural networks, whose convergence proofs\cite{Xu2024} are generally based on Lipschitz continuity. At first glance, this network is not Lipschitz continuous, as our nonlinearity is squaring, which has unbounded derivative on the real line. However, the inputs to each layer can empirically be seen to be within $\epsilon \approx 0.5$ of the initial and final range $[0,1]$, which is expected as each layer is only a slight alteration of the initial spectral refinement technique, which is formally bound within that range. This ensures Lipschitz continuity, but not necessarily in the same manner as such networks. We observe far stronger convergence behavior in layer count, and much faster convergence in training by using methods designed for curve-fitting tasks: rootfinding. Because we have a weight initialization (the embedding of traditional SP2) already very close to a solution, rootfinding methods such as nonlinear least squares\cite{Transtrum2012} display rapid convergence with such steps.

Although we see some key distinctions between traditional deep learning approaches and our ``Spectral Refinement Networks," many principles of deep learning still apply to our models, especially concerning expressivity. We show that using fully connected layers in this architecture appreciably improves model accuracy vs layer count, seen in Fig. \ref{fig:accuracy vs layers}, very similarly to how it does in CNNs.\cite{Huang2017} More generally, parsimoniously introducing skip connections as in ResNet\cite{He2015} can substantially improve the expressibility of the model without adding as severe memory requirements as full connectivity, while maintaining such theoretical benefits as smoothing the loss landscape.\cite{Hao2018}
Thus, drawing inspiration from machine learning, one can construct several families of such spectral networks, with increasing numbers of parameters (but also increasing memory requirements) that increase the accuracy of approximation at a given layer count.

\section{Model Embeddings \\ (SP2 $\subset$ MLSP2 $\subset$ Skip-SP2 $\subset$ Max-SP2 $\subset$ Arb-SP2)}
\label{appendix:embeddings}

Each of the following correspondences is quite useful not only analytically in showing heirarchies of expressibility, but also numerically, as it provides stable initial guesses, test cases, and benchmarks of errors introduced by changing the algorithm without changing the corresponding weights. Initial guesses are, in particular, extremely important, as they are minimally changed by the optimization process but lead to extreme gains in accuracy depending on the number of parameters available for fine tuning. We do not know whether different initial guesses from a less expressive model cause a more expressive model to fall into different basins of attraction in parameter space, and these relations can prove extremely useful in investigating that.

Obviously, we can show SP2 $\subset$ MLSP2 with the relation
\begin{equation} \begin{split}
    \theta_{i,1...4} &= \left\{ \begin{matrix}
    \{ 1, 0, 0, 0\} & \text{if } |\mu_i^2 - \mu' | < |2\mu_i^2 - \mu_i^2 - \mu'| \\ 
    \{-1, 2, 0, 0\} & \text{otherwise,}
\end{matrix} \right.
\end{split} \end{equation}
Showing MLSP2 $\subset$ Skip-SP2 is perhaps the most nuanced, as it involves going from an arbitrary quadratic to squaring an affine combination, which means certain terms must be shifted to the accumulator, and the linear combination must have an added constant term (hence, affine). Let's redefine, for convenience, MLSP2 as
\begin{equation} \begin{split}
    x_{i+1} &= a_i x_{i}^2 + b_i x_{i} + c_i, \\ 
    A &= \sum_{i=0}^{n} d_i x_i \; .
\end{split} \end{equation}
We wish to put it into the form
\begin{equation} \begin{split}
    \tilde x_{i+1} &= \left( \alpha_{i,k} + \alpha_{i,0} \; \tilde x_i + \beta_{i,1} A_{i,1}  \right)^2, \\ 
    A &= A_0 + \sum_{i=0}^{n} \gamma_i \tilde x_{i} \; ,
\end{split} \end{equation}
using $k=1$ and $K=1$. It's clear that $\beta_{i,j}=0$, and indeed, if we have
\begin{equation}
    x_0 = \tilde x_0 = 1 - x \; ,
\end{equation}
we can define
\begin{equation}
    \tilde x_{i+1} = \left( x_{i} + \frac{b_i}{2a_i} \right)^2 = x_{i}^2 + \frac{b_i}{a_i} \; x_{i}  + \frac{b_i^2}{4a_i^2} \; .
\end{equation} 
Then, $x_{i+1}$ can be put in terms of $\tilde x_{i+1}$:
\begin{equation}
    x_{i+1} = a_i \tilde x_{i+1} -\frac{b_i^2}{4a_i} + c_i  \; .
\end{equation} 
This definition is, of course, doubly recursive, but we can compactify it:
\begin{equation}
    \tilde x_{i+1} = \left( a_{i-1} \; \tilde x_{i} - \frac{b_{i-1}^2}{4a_{i-1}} + c_{i-1} + \frac{b_{i}}{2a_i} \right)^2 \; .
\end{equation} 
Thus, we have
\begin{equation} \begin{split}
    \alpha_{i,k} &= -\frac{b_{i-1}^2}{4a_{i-1}} + c_{i-1} + \frac{b_i}{2a_i} \\
    \alpha_{i,0} &= a_{i-1} \; .
\end{split} \end{equation}
Similarly, we have
\begin{equation} \begin{split}
    A &= A_0 + \sum_{i=0}^{N} d_i x_{i} \\ 
    &= A_0 + \sum_i d_i \left( a_{i-1} \; \tilde x_{i} -\frac{b_{i-1}^2}{4a_{i-1}} + c_{i-1} \right)  \; .
\end{split} \end{equation}
This gives
\begin{equation} \begin{split}
    \gamma_i &= d_i \; a_{i-1} \\
    A_0 &= \sum_j d_j \left( c_{j-1} - \frac{b_{j-1}^2}{4a_{j-1}} \right) \\
\end{split} \end{equation}
The alternative form of MLSP2, Eq.~(\ref{eq:mlsp2_alt}), is easiest to show from the above embedding in Skip-SP2. We require simply to factor out $\alpha_{i,0}$ from every intermediate expression. Indeed, we have the embedding in Skip-SP2
\begin{equation}\begin{split}
    \tilde x_{i+1} &= \left( \alpha_{i,k} + \alpha_{i,0} \; \tilde x_i \right)^2 = \alpha_{i,0}^2 \left( \frac{\alpha_{i,k}}{\alpha_{i,0}} + \tilde x_i \right)^2 \;.
\end{split}\end{equation}
If we allow every layer $x_i$ of MLSP2 to vary from $\tilde x_i$ by some multiplicative factor, i.e. $ s_i x_i = \tilde x_i$, then we can write
\begin{equation}\begin{split}
    s_{i+1} \; x_{i+1} &= \alpha_{i,0}^2 \left( \frac{\alpha_{i,k}}{\alpha_{i,0}} + s_i \; x_i \right)^2 \\
    &= \alpha_{i,0}^2 \; s_i^2 \left( \frac{\alpha_{i,k}}{\alpha_{i,0} \; s_i} + \; x_i \right)^2 \;.
\end{split}\end{equation}
By setting $s_{i+1} = \alpha_{i,0}^2 \; s_i^2$, we see that $\theta_{i,1} = \alpha_{i,k}/(\alpha_{i,0} s_i)$ and $\theta_{i,2} = \gamma_i s_i$ satisfy both the above and Eq.~(\ref{eq:mlsp2_alt}):
\begin{equation}\begin{split}
    x_{i+1} &= \left( \frac{\alpha_{i,k}}{\alpha_{i,0} \; s_i} + \; x_i \right)^2 \\
    A_{i+1} &= A_i + \gamma_i s_i x_i \; .
\end{split}\end{equation}
The final loose ends are $\theta_{n,1} = A_0$ and $\theta_{n,2} = \gamma_n s_n$.
Skip-SP2 $\subset$ Max-SP2 is given by
\begin{equation} \begin{split}
    \delta_i &= \left( \sum_{j=1}^K \beta_{i,j} A_{0,j} \right) +\alpha_{i,k} \\
    \theta_{i,j} &= \left( \sum_{l=1}^K \beta_{i,l} \gamma_{j,l} \right) + \left\{ \begin{matrix}
        \alpha_{i,n-j} & \text{ if } j \geq n-k \\
        0 & \text{otherwise.}
    \end{matrix} \right. \\
\end{split} \end{equation}
Finally, Max-SP2 $\subset$ Arb-SP2 is trivial, and has an additional degree of freedom for every layer: $c_{i} \psi_{i,j} = c_{i}^{-1} \phi_{i,j} = \theta_{i,j}$, where $c_i$ are arbitrary scalars that can be introduced to break symmetry.

\subsection{Eigenvalues of $\phi_i \psi_i^\top$ }

Each layer of Arb-SP2 as described by Eq.~(\ref{eq:arbitrary}) can be considered a quadratic form $\phi_i \psi_i^\top$ in the vector of all previous layer outputs, $x_{<i}$. To be specific, when operating on scalars, $\vec x_{<i} \in \mathbb R^i$, and when operating on matrices $x_{<i}$ is a module over the (abelian) ring of matrix polynomials in $H$, but the analysis proceeds identically as in the case of real numbers, so we shall consider it as such. The expressivity of this architecture is tied intrinsically to the rank of this quadratic form, which we examine below. 

The rank-one matrix $\mathbf u \mathbf v^\top$ has one (right) eigenvector $\hat {\mathbf u}$, with $\lambda = \mathbf u \cdot \mathbf v$, but when it is considered as a quadratic form, only its symmetric part, $\frac{1}{2}(\mathbf u \mathbf v^\top + \mathbf v \mathbf u^\top)$ acts on the input (the input is applied from both the left and the right, and thus the antisymmetric component $\mathbf x^\top \frac{1}{2}(\mathbf u \mathbf v^\top - \mathbf v \mathbf u^\top) \mathbf x$ cancels). This component, however, has two eigenvalues, which can be found by applying a similarity transformation that puts the matrix in block-diagonal form, in the basis of $\{\mathbf u, \mathbf v\}$ and additional directions that map to 0. In this basis, the nonzero block is
\begin{equation}
    M = \frac{1}{2} \begin{bmatrix}
    \langle \mathbf v, \mathbf u \rangle & \langle \mathbf v, \mathbf v \rangle \\
    \langle \mathbf u, \mathbf u \rangle & \langle \mathbf u, \mathbf v \rangle
\end{bmatrix} \; ,
\end{equation}
whose eigenvalues can be found by $\text{tr} M = \langle \mathbf u, \mathbf v \rangle $ and $\det M = \frac{1}{4} \left( \langle \mathbf u, \mathbf v \rangle^2 - \langle \mathbf u, \mathbf u \rangle \langle \mathbf v, \mathbf v \rangle \right)$. This gives
\begin{equation} \begin{split}
    \lambda &= \frac{\text{tr}M \pm \sqrt{(\text{tr} M)^2 - 4 \det M}}{2} \\
    &= \frac{1}{2}\left( \langle \mathbf u, \mathbf v \rangle \pm \sqrt{\langle \mathbf u, \mathbf u \rangle \langle \mathbf v, \mathbf v \rangle}\right) \;,
\end{split} \end{equation}
which are necessarily the same as the eigenvalues of the original matrix, because similarity transformations preserve eigenvalues. 

Thus, despite requiring only one general matrix-matrix multiplication, $\phi_n \psi_n^\top$ as a quadratic form is rank 2, and thus cannot be reduced to a single squaring (rank 1). This nearly doubles expressivity versus layer count, but requires arbitrary multiplication and not squaring -- a more expensive nonlinearity on specialized hardware. On Nvidia devices making full use of mixed precision arithmetic, matrix multiplications of this type are 50\% more expensive than squaring symmetric matrices on a flop basis, which can be seen by adapting the algorithm in Section \ref{sec:tensorcores} for the multiplication of two matrices $A,B$. This therefore sacrifices polynomial order achievable with a fixed number of hardware multiplications, which nonetheless may be useful at low $\beta$ (high temperature systems), where the cutoff is not so steep, and thus the dominant factor in the approximation accuracy is not simply polynomial order.

\subsection{Symmetry Breaking}
Propagating gradients through a layer of equation \ref{eq:arbitrary}  results in
\begin{equation} \begin{split} \label{eq:arbgrad}
\bar \phi_{i,j} &= \bar x_{i+1} \; \frac{\partial x_{i+1}}{\partial \phi_{i,j}} \\
&= \bar x_{i+1} \left(  \sum_{l \leq i} \psi_{i,l} x_l \right) x_j \; .
\end{split} \end{equation} 
Likewise for $\psi_i$,
\begin{equation} \begin{split} \label{eq:arbgrad2}
\bar \psi_{i,j} &= \bar x_{i+1} \left(  \sum_{l \leq i} \phi_{i,l} x_l \right) x_j \; .
\end{split} \end{equation} 
If the weights for Arbitrary Multiplication are naively initialized as the embedding of Max-SP2, i.e. $\phi_i = \psi_i = \theta_i$, then we see a critical issue. Equations \ref{eq:arbgrad} and \ref{eq:arbgrad2} become equal, so any optimization performed on $\phi, \psi$ will continue to keep them equal. However, the embedding of Max-SP2 into Arbitrary Multiplication has a particular ``gauge freedom." The rescaling $c_i \phi_i = c_i^{-1} \psi_i = \theta_i$ results in an identical model, as the constant factors can be factored out of the sums and cancel upon multiplication of the resultants. This cancellation ensures the first iteration will have identical $\frac{\partial x_{m>n}}{\partial x_{n+1}}$, and thus identical $\bar x_{n+1}$, but fundamentally differ in the second term. Comparing to the gradients in Max-SP2, we can reduce \ref{eq:arbgrad}-\ref{eq:arbgrad2} to
\begin{equation} \begin{split} \label{eq:arbgrad3}
\bar \theta_{i,j} &= 2 \bar x_{i+1} \left(  \sum_{l \leq i} \theta_{i,l} x_l \right) x_j \\
c_i^{-1} \bar \phi_{i,j} &= \tfrac{1}{2} \bar \theta_{i,j} \\
c_i \bar \psi_{i,j} &= \tfrac{1}{2} \bar \theta_{i,j} \; ,
\end{split} \end{equation} 
which breaks our symmetry by causing $\phi$ and $\psi$ to evolve at rates inversely proportional to their initial proportionality to $\theta$. Note that if $c_i \neq 1$, equation \ref{eq:arbgrad3} is only valid for the first step, after which symmetry breaks and $\phi_i, \psi_i$ are no longer proportional to $\theta_i$.

\section{Quadratic Convergence of SP2 when using repeated applications of $f_1 \circ f_0$}

\noindent If $0<x<\varphi^{-1}$, and $F=f_1\circ f_0$,
\begin{equation} \begin{split}
&| F(x) - \Theta(\varphi^{-1} - x) | \\
&=|F(0)+F'(0)x + \tfrac{1}{2} F''(0)x^2 -0 | \\
&= 2x^2 \; ,
\end{split} \end{equation}
so that if $x_n = F \circ \cdots \circ F(x)$ is $n$-th application of $F$ at $x$,
\begin{equation} \begin{split}
    | x_{n+1} - \Theta(\varphi^{-1} - x)| 
    & = | F(x_n) - \Theta(\varphi^{-1} - x)| \\
    & = 2(x_n - \Theta(\varphi^{-1} - x) )^2 \; ,
\end{split} \end{equation}
and $x_n \to 0$ with second order. A similar Taylor expansion about $x=1$ shows second order convergence of $x_n \to 1$ when $\varphi^{-1} < x < 1$. 

\vfill
\pagebreak


\end{document}